\documentclass{article}
\usepackage{arxiv}
\usepackage{appendix}
\usepackage{cite}
\usepackage{amsmath,amssymb,amsfonts}
\usepackage{graphicx}
\usepackage{textcomp}
\usepackage{amsthm}
\usepackage{enumitem}
\usepackage{algorithm}
\usepackage{algpseudocode}
\usepackage{xurl} 
\usepackage{color}
\usepackage{xcolor}
\usepackage{soul}
\sethlcolor{yellow}
\usepackage[hidelinks]{hyperref}
\hypersetup{breaklinks=true}
\usepackage[caption=false,font=normalsize,labelfont=sf,textfont=sf]{subfig}
\newtheorem{definition}{Definition}
\newtheorem{theorem}{Theorem}
\newtheorem{proposition}{Proposition}
\newtheorem{corollary}{Corollary}

\usepackage{bm}
\makeatletter
\AtBeginDocument{\DeclareMathVersion{bold}
\SetSymbolFont{operators}{bold}{T1}{times}{b}{n}
\SetMathAlphabet{\mathrm}{bold}{T1}{times}{b}{n}
\SetMathAlphabet{\mathit}{bold}{T1}{times}{b}{it}
\SetMathAlphabet{\mathbf}{bold}{T1}{times}{b}{n}
\SetMathAlphabet{\mathtt}{bold}{OT1}{pcr}{b}{n}
\SetSymbolFont{symbols}{bold}{OMS}{cmsy}{b}{n}
\renewcommand\boldmath{\@nomath\boldmath\mathversion{bold}}}
\makeatother

\def\BibTeX{{\rm B\kern-.05em{\sc i\kern-.025em b}\kern-.08em
    T\kern-.1667em\lower.7ex\hbox{E}\kern-.125emX}}

\title{Incentive Mechanism Design for Privacy-Preserving Decentralized Blockchain Relayers}

\author{
Boutaina Jebari$^{1,2,3}$ \thanks{Corresponding author: \texttt{boutaina.jebari@uit.ac.ma}}  \quad
\textbf{Khalil Ibrahimi}$^{1}$ \quad
\textbf{Hamidou Tembine}$^{4,5}$ \quad
\textbf{Mounir Ghogho} $^{2,6}$  \\[2ex]
$^{1}$Ibn Tofail University, Faculty of Sciences, LaRI Laboratory, Morocco \\ \\
$^{2}$ University Mohammed VI Polytechnic, College of Computing, Morocco \\ \\
$^{3}$ International University of Rabat,  TICLab Laboratory, Morocco \\ \\
$^{4}$ Department of Electrical and Computer Engineering, School of Engineering, \\
University of Quebec at Trois-Rivieres, Quebec, Canada \\ \\
$^{5}$ Learning and Game Theory Laboratory, TIMADIE, Paris, France \\ \\
$^{6}$ University of Leeds, Faculty of Engineering, UK \\ [1ex]
\texttt{boutaina.jebari@uit.ac.ma}, \texttt{ibrahimi.khalil@uit.ac.ma} \\  \texttt{hamidou.tembine@uqtr.ca}, \texttt{mounir.ghogho@um6p.ma} 
}

\begin{document}
\maketitle
\begin{abstract}
Public blockchains, though renowned for their transparency and immutability, suffer from significant privacy concerns. Network-level analysis and long-term observation of publicly available transactions can often be used to infer user identities. To mitigate this, several blockchain applications rely on relayers, which serve as intermediary nodes between users and smart contracts deployed on the blockchain. However, dependence on a single relayer not only creates a single point of failure but also introduces exploitable vulnerabilities that weaken the system’s privacy guarantees. This paper proposes a decentralized relayer architecture that enhances privacy and reliability through game-theoretic incentive design. We model the interaction among relayers as a non-cooperative game and design an incentive mechanism in which probabilistic uploading emerges as a unique mixed Nash equilibrium. Using evolutionary game analysis, we demonstrate the equilibrium’s stability against perturbations and coordinated deviations. Through numerical evaluations, we analyze how equilibrium strategies and system behavior evolve with key parameters such as the number of relayers, upload costs, rewards, and penalties. In particular, we show that even with high transaction costs, the system maintains reliability with an outage probability below $0.05$. Furthermore, our results highlight a fundamental trade-off between privacy, reliability, robustness, and cost in decentralized relayer systems.
\end{abstract}


\section{Introduction}
Public blockchains introduce a new design paradigm for digital infrastructures by providing user-controlled, transparent, and tamper-resistant execution environments. Through decentralized smart contracts and publicly verifiable ledgers, they enable applications ranging from decentralized finance to supply chain management, voting, and open data markets. However, public blockchains also raise significant privacy concerns. Since all interactions with smart contracts are permanently and publicly visible, adversaries can perform network-level analyses and long-term observations to extract sensitive information about users. Such analyses can link off-chain identities to on-chain behavior, thereby undermining user privacy \cite{survey_privacy,deanonymization}.

\par To address these privacy challenges, several approaches have been proposed in the literature. A first approach relies on advanced cryptographic primitives, such as zero-knowledge proofs and threshold credential schemes like Coconut, to enable privacy-preserving authentication without revealing sensitive information \cite{zklogin,nymCredentials}. A second category focuses on network-level anonymity, where protocols such as Dandelion and random walks improve source untraceability through probabilistic message dissemination \cite{guerraouiPrivacy}. More recently, hybrid architectures that combine layered routing principles with blockchain-based incentive mechanism have emerged. Notably, the Nym network implements an incentivized mixnet where nodes are rewarded for contributing to privacy-preserving communication \cite{nymNetwork}.  Additionally, for smart contract interactions, blockchain systems have introduced the concept of relayers, intermediary nodes that submit transactions to the blockchain on behalf of users \cite{relayers_def}. Relayers decouple the user’s on-chain identity from the origin of the transaction, thereby enhancing privacy. However, most existing implementations rely on a single relayer, which introduces new vulnerabilities. A single relayer becomes a point of failure: an adversary controlling or monitoring this node can exploit timing, traffic correlation, or propagation information to infer the transaction’s source \cite{flexible_privacy,active_passive}. In practice, this reduces the level of anonymity and exposes users to targeted attacks.
\par
While relayers improve transaction privacy by decoupling users from their on-chain activity, relying on a single intermediary introduces a single point of failure and exposes users to potential deanonymization attacks. A natural extension is to distribute this role among multiple relayers, allowing several independent nodes to participate in the transaction forwarding process. However, this decentralization introduces a new coordination problem: while involving multiple relayers enhances privacy by enlarging the anonymity set, it can also reduce efficiency. Moreover, without proper incentives, rational nodes may either rush to relay, revealing information about message propagation, or refrain from participating to avoid costs, undermining reliability. Hence, the challenge lies in designing an incentive mechanism that enables decentralized coordination among relayers, ensuring that at least one relayer transmits to the  blockchain, while preserving privacy and fairness across participants.
\par
In this work, we propose an incentive mechanism for decentralized blockchain relayers designed to improve transaction privacy. Relayer participation is modeled as a strategic decision problem, and the resulting equilibrium behavior is analyzed. The main contributions of this work are summarized as follows:
\begin{itemize}
    \item We introduce a game-theoretic framework for decentralized relayers in which the incentive mechanism enforces indistinguishable probabilistic actions. At equilibrium, all nodes adopt the same upload strategy, which is required to preserve privacy against on-chain observers.

    \item We model the relayer interaction as a variant of the Volunteer's Dilemma (VD) with symmetric rewards and a probabilistic cost that reflects accepted and reverted smart-contract submissions, together with a penalty when no relayer participates.

    \item We establish the existence, uniqueness, and evolutionary stability of the symmetric mixed-strategy Nash equilibrium (NE) of the game.

    \item We analyze the privacy–reliability–robustness–cost trade-off induced by the design and use numerical evaluations to illustrate how equilibrium behavior varies with system parameters.
\end{itemize}
The remainder of this paper is organized as follows. Section II reviews related work, Section III presents the system model and problem formulation, Section IV introduces the game-theoretic model and provides the analytical results on equilibrium existence, stability, and robustness. Section V presents the numerical evaluations and discusses the observed trade-offs between privacy, reliability, and robustness. Finally, Section VI concludes the paper and highlights directions for future research.

\section{Related Work}

\subsection{Privacy in Public Blockchains}

Privacy in public blockchains has been addressed through several complementary approaches that operate at different layers. Cryptographic mechanisms, such as zero-knowledge proofs and threshold credential systems, provide strong unlinkability guarantees at the application or data layer but introduce significant computational overhead \cite{zerocash2014,nymCredentials}. Network-layer anonymity mechanisms aim to obscure transaction origin during propagation; protocols such as Dandelion++ and its subsequent analyses rely on controlled gossiping and probabilistic routing to reduce linkability
\cite{dandelionpp,goncalves2022dandelionSecurity}.
More recent designs combine cryptographic techniques with layered routing, as in mixnet-based architectures such as Loopix, where batching and cover traffic further strengthen sender anonymity
\cite{loopix}.
While each of these approaches mitigates specific privacy risks, none is sufficient on its own, and practical systems often combine techniques across layers to address complementary threats.
\par
Relayers represent another practical approach to improve transaction privacy by decoupling a user’s network identity from the submitted transaction. They appear in interoperability systems such as zkRelay \cite{zkRelay} and in Ethereum’s account abstraction framework, where bundlers submit transactions on behalf of users \cite{EIP4337}. Relayers are also used in operational infrastructures such as OpenZeppelin Defender \cite{openzeppelin_relayers} and in privacy-preserving mixers such as Tornado Cash \cite{tornado}, where decentralized relayer sets help prevent linkability between deposits and withdrawals. In these designs, the privacy threat is an adversary observing on-chain interactions with smart contracts. These systems, however, treat relayer behavior as static and trusted.
\par
Incentive mechanisms have been used in several blockchain settings, including consensus protocols and oracle networks, to align participant behavior through rewards and penalties. The closest system to an incentive-driven privacy mechanism is Nym, an incentivized mixnet where nodes are rewarded for forwarding packets \cite{nymNetwork}. While Nym introduces incentives to support network-layer anonymity, existing relayer-based architectures do not incorporate incentives designed to enforce indistinguishable behavior among relayers. As a result, privacy depends on honest operation rather than on a principled, incentive-compatible design.

\subsection{Volunteer's Dilemma}

The classical volunteer’s dilemma introduced in \cite{volunteersDilemma85} describes settings where a single participant’s action is sufficient for all participants to gain a reward. If at least one agent volunteers, all agents receive the same payoff, while the volunteer incurs a cost. If no one does, no participant gains any benefit. The analysis presented in \cite{volunteersDilemma85} shows the existence of a unique symmetric mixed-strategy Nash equilibrium that determines each player’s volunteering probability. It also highlights how this probability declines as the number of participants grows, illustrating the familiar free-riding effect that tends to arise in such situations.

\par
Another work \cite{myatt_oxford_2008} considers an asymmetric version of the volunteer’s dilemma in which players differ in their willingness to volunteer and in their likelihood of maintaining the volunteering action over time. The authors study the evolutionary dynamics of the game and show that long-run outcomes do not necessarily favor the lowest-cost individual. Instead, the eventual volunteer tends to be the agent who is both inclined to take action and unlikely to withdraw once doing so. Their framework models behavioral adjustments through a birth–death process in which players may switch into or out of volunteering. Similarly, the authors in \cite{konrad} study the volunteer’s dilemma in finite populations in which players differ in the probability of volunteering. The work examines which volunteering probabilities persist under evolutionary dynamics. Their analysis shows that, in pairwise population interactions, types that volunteer less often have a relative evolutionary advantage, although populations consisting entirely of more cooperative types can still appear in the long-run distribution. They further show that higher rewards and smaller groups make cooperative behavior more likely whereas in sufficiently large groups evolutionary pressures lead almost exclusively to types that volunteer infrequently, which is consistent with the symmetric case.

\par
Variants of the volunteer’s dilemma have also been applied to communication networks. In particular, several works model probabilistic forwarding in vehicular ad hoc networks as a volunteer’s-dilemma interaction \cite{vd_naja, vd_vanet}. These models capture the tradeoff between message dissemination and communication overhead, and show how VD-type incentives regulate decentralized participation.
\par
To the best of our knowledge, prior volunteer’s-dilemma formulations have not been applied to blockchain relayers or to smart-contract transaction submission. Existing relayer infrastructures typically treat relayer behavior as fixed and do not introduce incentive mechanisms that enforce indistinguishable probabilistic actions. In addition, classical VD models do not capture two elements that arise when a set of relayers interacts with a smart contract: the realized volunteering cost depends on whether the transaction is accepted or reverted, which is not deterministic, and staked participation enables the protocol to impose a penalty when no participation occurs. These aspects differ from the assumptions commonly used in the VD literature and motivate the formulation adopted in this work, which is presented in detail in the next section.

\section{Problem Formulation}

We consider the privacy of users interacting with a blockchain through a relayer system, focusing on an observer who monitors on-chain transactions and attempts to infer the origin of submitted messages. The observer is assumed to have access only to blockchain-level information, such as transaction contents, the relayer address that submitted them, and the time of inclusion. We do not consider network-layer traffic analysis or cryptographic anonymity mechanisms and focus instead on indistinguishability at the blockchain layer.
\par
We begin with the centralized case, in which a single relayer submits transactions on behalf of all users, this action is called an upload. Here, the observer cannot distinguish which user initiated any particular transaction, since only the relayer’s address appears on-chain. Although this provides a basic form of privacy, it introduces a single point of failure and requires users to trust the relayer for correctness, availability, and fairness.
\par
To remove this trust assumption, we turn to a decentralized set of relayers. Users' transactions are  disseminated to all relayers, and any one of them may upload the transaction to the smart contract. We abstract away the network-level dissemination protocol and any cryptographic mechanisms that ensure transaction validity and focus solely on the blockchain-facing behavior relevant to our incentive design approach to privacy. A user choosing to act as a relayer becomes indistinguishable from the other relayers, while a user who does not participate remains indistinguishable among all non-relayer users. This decentralized structure enlarges the anonymity set and avoids centralized trust, but it introduces coordination challenges among relayers.
\par
If only the successful uploader is rewarded, rational relayers will race to upload as quickly as possible. In practice, the relayer that first receives the message is most likely to upload, which reveals information about message propagation and can degrade privacy. Such incentives may also encourage relayers to interfere with competitors to increase their chances of uploading. The central challenge is therefore to design an incentive mechanism that encourages behavior that renders all relayers indistinguishable from the perspective of the observer.

\subsection{System and Model Assumptions}

We consider a decentralized set of off-chain relayers, each of whom stakes assets to participate in the system. Staking provides Sybil resistance and allows the protocol to apply penalties when relayers fail to perform their role. All relayers operate under the same protocol-defined rules, which ensures that their on-chain behavior is uniform and supports the indistinguishability requirement introduced in the privacy model.
\par
Each user message is disseminated to all relayers before an upload window begins. The upload window is derived from the blockchain state, for example from the block height or slot number, so all relayers observe the same interval. This avoids dependence on local clocks and prevents timing differences that could reveal which relayer received the message first. During the upload phase, each relayer decides whether to upload the message to the smart contract. The cost of uploading depends on the outcome: the successful uploader pays a cost $c_f$, while unsuccessful attempts incur a lower cost $c_l$ with $0 < c_l < c_f$. This lower cost represents the transaction cost of a reverted transaction.  Since transactions broadcast within the same block interval face essentially the same gas price, assuming symmetric cost parameters is consistent with blockchain execution. A successful upload yields a reward $b > c_f$, while failure to upload before the deadline results in a penalty $p > 0$ applied to all relayers. Both the reward and the penalty are enforced through the staking contract. Our analysis concerns a single upload window; the dynamic fees resulting from repeated interactions are outside the scope of our model and are discussed in the discussion section.
\par
If the protocol were to reward only the first uploader, rational relayers would race to upload as quickly as possible. This reveals information about message propagation and undermines privacy. Pre-selecting a single uploader through public randomness would avoid contention but requires decentralized and unpredictable randomness, which remains difficult to achieve in practice. Existing mechanisms such as RANDAO are vulnerable to prediction, while verifiable delay functions and threshold beacons introduce significant operational overhead.
\par
\begin{figure}[htbp]
\begin{center}
\includegraphics[width=0.9\textwidth]{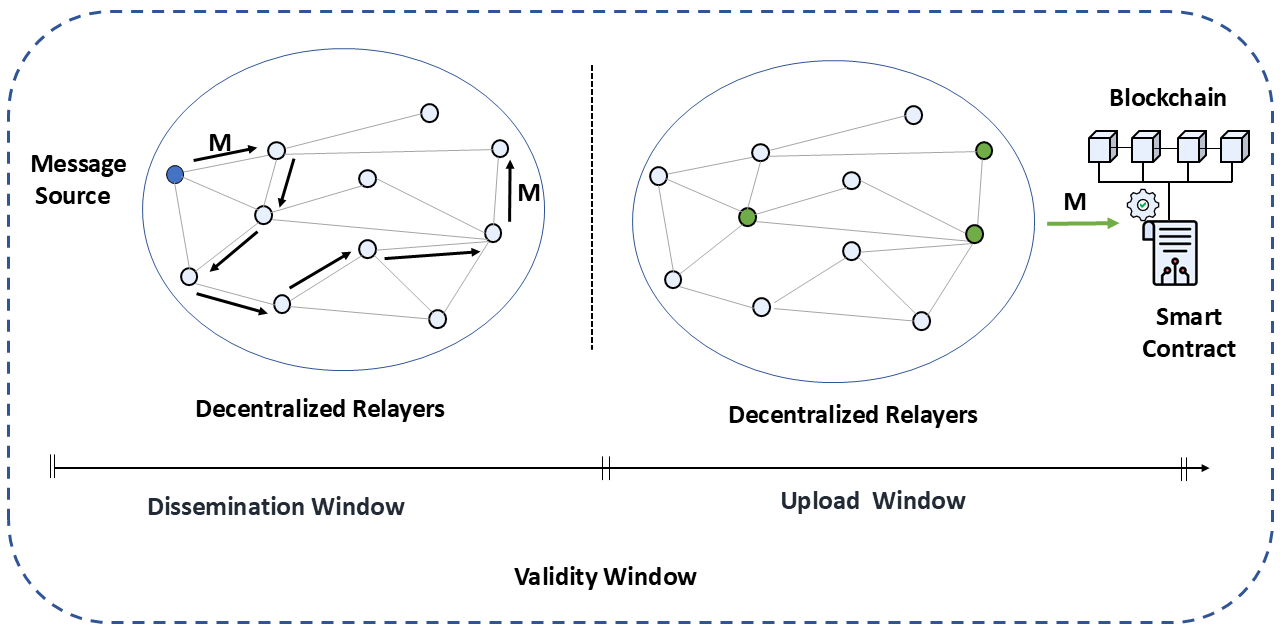}
\caption{Illustration of the decentralized relayer model.}
\label{illustration}
\end{center}
\end{figure}

For these reasons, we adopt a model in which relayers attempt uploads independently and with the same probability. From the perspective of an observer limited to blockchain-level information, each relayer is therefore an equally plausible uploader. This yields an anonymity set whose size equals the number of participating relayers and prevents information leakage from deterministic or propagation-ordered behavior. The overall system architecture is illustrated in Figure~\ref{illustration}.

We analyze the decentralized relayer system under the following assumptions:

\begin{itemize}
    \item The system contains a fixed set of $N \geq 3$ staked relayers.
    \item All relayers receive the disseminated message before the upload window begins.
    \item The upload window is derived from blockchain state, so all relayers observe the same timing.
    \item Each relayer chooses between two actions during the upload phase: \emph{Upload} or \emph{Not Upload}.
    \item Each relayer uploads independently with probability $q_u$.
    \item If multiple uploads occur, the protocol selects the first valid transaction uniformly at random. The first one is accepted and the rest are reverted. 
    \item Upload costs are outcome-dependent: the successful uploader pays $c_f$ and unsuccessful attempts pay $c_l$ with $0 < c_l < c_f$.
    \item A successful upload yields a reward $b > c_f$; failure to upload before the deadline triggers a penalty $p > 0$ applied to all relayers.
    \item All parameters $N, b, c_f, c_l, p$ are public and enforced by the smart contract.
    \item The external observer is limited to blockchain-level information and does not observe network-layer events.
\end{itemize}

To illustrate a possible use case, consider privacy-preserving incentivized data collection, where contributors must later submit a commitment proving that they provided data. Uploading these commitments directly would reveal their activity. By disseminating the commitment to a set of relayers, one of whom uploads it on-chain, the contributor becomes indistinguishable from the other relayers and thereby preserves privacy.

\section{The Relayer Upload Game}
In this section, we formally model the decentralized relayer system as a symmetric non-cooperative game and analyze the existence, uniqueness, and stability of its Nash equilibrium.
\subsection{The Game Model }

We formally model the upload process as a finite game, denoted as $\mathcal{G}_N$, where $\mathcal{G}_N = \langle \mathcal{N}, \mathcal{A}, (u_i)_{i \in \mathcal{N}} \rangle$.

\par
\textit{Players $\mathcal{N}$:} The players in the game correspond to the nodes in our peer-to-peer relayer network. We model them as a finite set $\mathcal{N} = \{1, ..., N\}$, where $N \in \mathbb{N}$ represents the total number of participating nodes and $N\geq3$.

\par
\textit{Action Space $\mathcal{A}$:} Each player decides whether or not to upload the message to the blockchain. Consequently, the action set for each player $i$ is given by $\mathcal{A}_i =\{\textrm{U}, \textrm{NU}\}$, where U denotes "Upload" and NU denotes "Not Upload." The general action space of the game is  $\mathcal{A} = \mathcal{A}_1 \times \mathcal{A}_2 \times \dots \times \mathcal{A}_N$. The action taken by player $i$ is denoted by $a_i \in \mathcal{A}_i$, and the joint action profile of all players is represented as $\underline{a} = (a_1, a_2, ..., a_N)$. 
\par 
\textit{Utility Function:}
The utility of each player $i \in \mathcal{N}$, denoted by $u_i(\underline{a})$, is defined in Equation \eqref{utilityEq}. In this setup, $b$ represents the reward distributed to the nodes, while $c_f$ and $c_l$ denote the costs of being the first to upload and of attempting to upload without being first, respectively.
\begin{equation}
u_i(\underline{a}) = 
\begin{cases}
b - c_f& \text{if } a_i = \textrm{U} \text{ and $i$ is first uploader } \\
b - c_l& \text{if } a_i = \textrm{U}\text{ and $i$ is not the first uploader } \\
b & \text{if } a_i = \textrm{NU} \text{ and someone else uploads}  \\
-p & \text{if no one uploads} 
\end{cases}
\label{utilityEq}
\end{equation}
Next, we analyze the equilibria of our relayer upload game. For the reader’s convenience, Table~\ref{tab:notation} provides a summary of the notation and parameters introduced in this section and used throughout the remainder of the analysis.
\begin{table}[t]
\caption{Notation Summary}
\label{tab:notation}
\centering
\begin{tabular}{p{0.22\linewidth} p{0.72\linewidth}}
\hline
\textbf{Symbol} & \textbf{Meaning} \\
\hline
$N$ & Number of relayers in the system, with $N \ge 3$. \\
$\mathcal{N}$ & Set of relayers (players), $\mathcal{N}=\{1,\dots,N\}$. \\
$i$ & Index of a relayer, $i \in \mathcal{N}$. \\
$\mathcal{A}_i$ & Action set of relayer $i$, $\mathcal{A}_i=\{\mathrm{U},\mathrm{NU}\}$. \\
$\mathcal{A}$ & Joint action space, $\mathcal{A}=\mathcal{A}_1 \times \cdots \times \mathcal{A}_N$. \\
$a_i$ & Action chosen by relayer $i$. \\
$\underline{a}$ & Joint action profile, $\underline{a}=(a_1,\dots,a_N)$. \\
$\mathrm{U}$ & Upload action. \\
$\mathrm{NU}$ & Not Upload action. \\
$u_i(\underline{a})$ & Utility of relayer $i$ under joint actions $\underline{a}$. \\
$b$ & Reward received by each relayer when an upload succeeds. \\
$c_f$ & Cost incurred by the first (successful) uploader. \\
$c_l$ & Cost incurred by an unsuccessful upload attempt, with $0<c_l<c_f$. \\
$p$ & Penalty incurred by each relayer when no upload occurs. \\
$q_u$ & Upload probability used in a symmetric mixed strategy. \\
$q_u^*$ & Symmetric mixed-strategy Nash equilibrium upload probability. \\
$v(\mathrm{U},q_u)$ & Expected utility of uploading, given others upload with probability $q_u$. \\
$v(\mathrm{NU},q_u)$ & Expected utility of not uploading, given others upload with probability $q_u$. \\
$g(q_u)$ & Expected gain from uploading: $g(q_u)=v(\mathrm{U},q_u)-v(\mathrm{NU},q_u)$. \\
$h(q_u)$ & Polynomial equilibrium condition whose root in $(0,1)$ equals $q_u^*$. \\
$P_O^*$ & Outage probability at equilibrium: $P_O^*=(1-q_u^*)^N$. \\
$R^*$ & Expected reward (utility) of a relayer at equilibrium. \\
$\phi(q_u)$ & Potential function of the symmetric game: $\phi(q_u)=\int_0^{q_u} g(x)\,dx + C$. \\
$\mu$ & Adaptation rate in the replicator dynamic. \\
$q_u(t)$ & Population upload probability at time $t$ under evolutionary dynamics. \\
$\dot{q}_u(t)$ & Time derivative of $q_u(t)$ in the replicator dynamic. \\
$q_m$ & Upload probability of a mutant strategy in evolutionary analysis. \\
$\epsilon$ & Fraction of mutants in the population. \\
$\alpha$ & Invasion barrier (maximum mutant fraction that cannot invade). \\
\hline
\end{tabular}
\end{table}
\subsection{Equilibrium Analysis for $\mathcal{G}_N$ }

In this part, we prove the existence and uniqueness of a symmetric mixed strategy Nash equilibrium and analyze its evolutionary stability. 

\par
We first recall the definition of the Nash equilibrium in Definition \ref{NE}.

\begin{definition}[Nash Equilibrium, \cite{SAMSON_CHAP1}]
\label{NE}
Let $\mathcal{S} = \mathcal{S}_1 \times \mathcal{S}_2 \times \dots \times \mathcal{S}_N$ and let $u_i : \mathcal{S} \xrightarrow{} \mathbb{R}, i \in \mathcal{N}$. The vector $\underline{s}^*$ is a Nash Equilibrium if: 
\[
\forall i \in \mathcal{N},\ \forall s_i \in \mathcal{S}_i, u_i(s_i^*, \underline{s}_{-i}^*) \geq u_i(s_i, \underline{s}_{-i}^*),
\]
where $\underline{s}_{-i}^* =  (s^*_1, ..., s^*_{i-1},s^*_{i+1},s^*_N )$.
\end{definition}
To analyze the existence and uniqueness of a symmetric mixed equilibrium, we define the function $g : [0,1] \to \mathbb{R} $ by:
\begin{equation}
    g(q_u) = v(\text{U}, q_u) - v(\text{NU}, q_u),
    \label{defg}
\end{equation}

which represents the difference in expected utility between uploading and not uploading, assuming all other players upload with probability $q_u$.  
A symmetric mixed Nash equilibrium corresponds to a point $q_u^* \in (0,1)$ such that $g(q_u^*) = 0$, i.e., where players are indifferent between the two actions.
\begin{theorem}[]
The game $\mathcal{G}_N$ has a unique symmetric NE where players upload with probability $q_u^*$, where $q_u^*$ is the unique root in the open interval $(0,1)$ of the following polynomial, 
\begin{equation*}
        h(q_u)  = Nq_u(b+p)(1-q_u)^{N-1} - N q_u c_l  + (c_l - c_f) (1-(1-q_u)^{N} ).
\end{equation*}
\label{uniquenessNE}
\end{theorem}
Consider the function
$g(q_u) = v(\mathrm{U}, q_u) - v(\mathrm{NU}, q_u),$
which represents the expected utility gain from uploading when all other relayers upload with probability $q_u$. When $q_u = 0$, no other relayer uploads, and a unilateral upload strictly improves the outcome, so $g(0) > 0$. When $q_u = 1$, all relayers upload, and uploading yields only its cost, implying $g(1) < 0$. Since $g$ is continuous, there exists at least one root in $(0,1)$. Substituting the explicit utilities yields the polynomial $h(q_u)$ given in Theorem~1. An analysis of its derivative shows that $h'(q_u)$ changes sign exactly once on $(0,1)$, so $h$ attains a single local maximum and crosses zero at a unique point. This point corresponds to the unique symmetric mixed-strategy Nash equilibrium $q_u^*$.  The proof is provided in Appendix~A.1. 
\par 
The existence of a unique mixed Nash equilibrium implies that when all nodes adopt a probabilistic upload strategy with probability $q_u^*$, the system reaches a single, well-defined equilibrium configuration. At this point, no individual node can unilaterally deviate and improve its expected payoff, and no alternative randomization profile gives a higher utility. 
\par 
Such equilibrium behavior arises when nodes are either initialized at the equilibrium configuration or possess prior knowledge of the equilibrium strategy. In practice, however, real systems are subject to deviations, whether intentional or unintentional. To assess the system’s resilience to such perturbations, we next examine its dynamic behavior under evolutionary dynamics. Before proceeding, we recall the definition of a potential game.  
\begin{definition}[Exact Potential Game, \cite{SAMSON_CHAP2}]
The game $\mathcal{G}$ is an exact potential game if there exists a function $\phi$ such that:
\begin{equation}
\begin{split}
    \forall i \in \mathcal{N},~\forall (s_i, \underline{s}_{-i}) \in \mathcal{S},~\forall s_i' \in \mathcal{S}_i, 
    u_i(s_i, \underline{s}_{-i}) - u_i(s_i', \underline{s}_{-i}) = \phi(s_i, \underline{s}_{-i}) - \phi(s_i', \underline{s}_{-i}).
\end{split}    
\end{equation}
\end{definition}

\begin{proposition}
The  game $\mathcal{G}_N$ is a potential game  with potential function : 
\begin{equation*}
    \phi(q_u) = \int_0^{q_u} g(x) \, dx + C, 
\end{equation*} 
where $g$ is defined in Equation \ref{defg} and $C \in \mathbb{R}$ is a constant. 
Moreover, the unique symmetric mixed NE $q_u^*$ is  asymptotically stable under any evolutionary dynamic $V_F$ that is Nash stationary and satisfies positive correlation.
\label{stability}
\end{proposition}
This result shows that the relayer upload game admits a potential function that captures  the collective behavior of all players. The Nash equilibrium corresponds to a stable point of this function, where any unilateral deviation reduces a player’s expected payoff. Moreover, the asymptotic stability property implies that when nodes temporarily deviate, the system naturally converges back to equilibrium, provided that players favor strategies with higher payoffs.
\begin{definition}[$\alpha$-Strong Equilibrium, \cite{SAMSON_CHAP1}]
\label{SNE}
A $\alpha$-strong (Aumann) equilibrium is a strategy profile \( s^* \) from which no coalition (of any size less or equal to $\alpha$) can deviate and improve the utility of every member of the coalition (denoted by \( \mathcal{C} \)), while possibly lowering the utility of players outside the coalition : 
\begin{equation}
\forall i \in \mathcal{N}, \forall \mathcal{C} \subset \mathcal{N}, \forall \underline{s}_{ \mathcal{C}}, \quad u_i(\underline{s}^*) \geq u_i( \underline{s}_{ \mathcal{C}}, \underline{s}^*_{ -\mathcal{C}}).
\label{eq:strong_eq}
\end{equation}
\end{definition}
%%%%

\begin{proposition}[$N-$strong equilibria of the relayer upload game] \label{propess1}
Consider the relayer upload game $\mathcal{G}_N.$ Then the set of $N-$strong equilibria of $\mathcal{G}_N$ is
  \begin{equation*}
      \begin{aligned}
          \mathrm{SE} &= \Bigl\{ \underline{a} \in \mathcal{A} : \text{exactly one player chooses U and all others} \\
                    & \qquad \text{ choose NU} \Bigr\}.
      \end{aligned}
  \end{equation*}

A strategy profile is a strong equilibrium if and only if exactly one node uploads the message to the blockchain while all remaining nodes choose not to upload.
\end{proposition}

\begin{corollary}
\label{col1}
Every pure strategy Nash equilibrium of this game is a global maximizer of $\sum_{i \in \mathcal{N}} u_i(s)$.
\end{corollary}

\begin{proposition}[Evolutionary Stability and Invasion Barrier] \label{propess2}
Consider the symmetric $N$-player relayer upload game. Let $q_u^* \in (0,1)$ denote the symmetric strategy used by almost all players (residents), and let $q_m \in (0,1)$ denote the strategy used by a small fraction of mutants. Then there exists a positive \emph{invasion barrier} $\alpha > 0$ such that if the fraction of mutants $\epsilon \le \alpha$, the resident strategy $q_u^*$ yields a strictly higher expected payoff than any mutant strategy $q_m \neq q_u^*$.
\end{proposition}

This result shows that restricted to symmetric strategies, the mixed symmetric Nash equilibrium is also a $\alpha$-strong equilibrium, meaning that no coalition of relayers can jointly deviate to improve their payoffs. It thus guarantees stability not only against individual deviations but also against coordinated group behavior.
\par 
We have established the existence, uniqueness, and stability of the equilibrium upload strategy. Next, we perform a numerical analysis to examine how the system behaves at equilibrium under different configuration parameters.
\section{Numerical Evaluation}
In this section, we present numerical results illustrating the system’s behavior at equilibrium and discuss the practical implications of these results.

\subsection{Evaluation Parameters and Metrics}
To evaluate the proposed model, we vary the main system parameters and measure their effect on the upload probability and overall system performance at equilibrium. Throughout all simulations, the benefit is fixed at $b = 100$, while the other parameters are varied as a ratio of the benefit. These parameters include: the cost of the first upload $c_f$, the cost of later uploads $c_l$ the penalty for a no-upload event $p$, and the number of relayers $N$. Since in practice, successful execution of method calls incurs higher cost than reverted ones, $c_l$ is chosen as a unit of cost. The ratio between $c_f$ and $c_l$ varies depending the actual code executed.  The numerical values are normalized  rather than literal gas measurements; only the ratios between $b, c_f, c_l$, and $p$ influence equilibrium behavior.
\par 
The number of relayers $N$ captures the degree of decentralization and quantifies the privacy level. The upload costs $c_f$ and $c_l$ represent the cost of interaction with the blockchain, mainly the price of the upload transaction, while the penalty $p$ acts as an incentive to prevent coordination failures.
\par
Across the evaluations, we focus on three key metrics to assess system performance: 

\begin{itemize}
    \item Equilibrium upload probability $q_u^*$: This represents the mixed-strategy equilibrium probability. It is obtained numerically by solving the equilibrium condition derived from the polynomial function $h$.
    \item Outage probability at equilibrium $P_O^*$: This denotes the probability that no node uploads during a round, when nodes play the equilibrium strategy. It represents the reliability of the system and is given by $P_O^* = (1 - q_u^*)^N$. A lower outage probability implies a more dependable system.
    \item Expected reward $R$: the equilibrium utility obtained by a node, which measures the degree of incentive compatibility. When $R$ remains positive even under partial failures, it demonstrates the robustness of the system’s incentive design. The expected reward at equilibrium is denoted $R^*$.
\end{itemize}

To further assess the system’s resilience, we analyze its stability under perturbations using the replicator dynamic, which models how the population of strategies evolves over time \cite{evolutionaryGameDynamic}. Particularly, for the symmetric case,  the replicator dynamic can be expressed as 
\begin{equation}
    \dot{q}_u(t) = \mu \, q_u(t)\, \big(1 - q_u(t)\big)\, \Big( v(\text{U}, q_u(t)) - v(\text{NU}, q_u(t)) \Big),
    \label{replicators_dynamic_GN}
\end{equation}
where $\mu > 0$ is the adaptation or convergence rate \cite{replicatorSymmetric}. 

\subsection{Numerical Results}
We evaluate the system’s equilibrium behavior by varying key parameters such as the number of relayers, the upload cost, and the penalty value. We show how these parameters influence the equilibrium upload probability, outage probability, players' payoff and overall system robustness.

\subsubsection{System Performances at Equilibrium}
\begin{figure}[htbp]
    \centering 
    \subfloat[]{% 
        \includegraphics[width=0.3\linewidth]{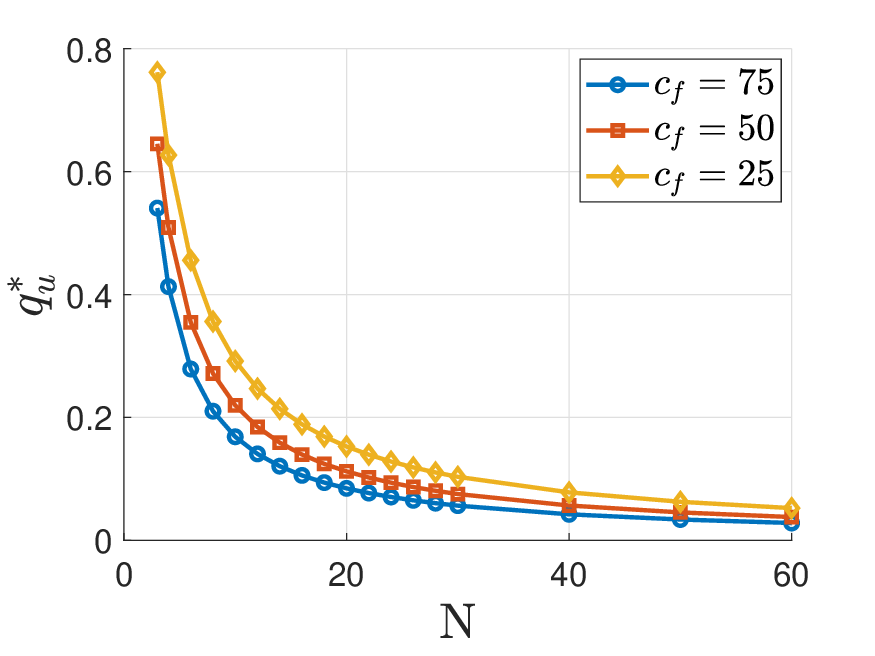}} 
     \hfill
    \subfloat[]{% 
        \includegraphics[width=0.3\linewidth]{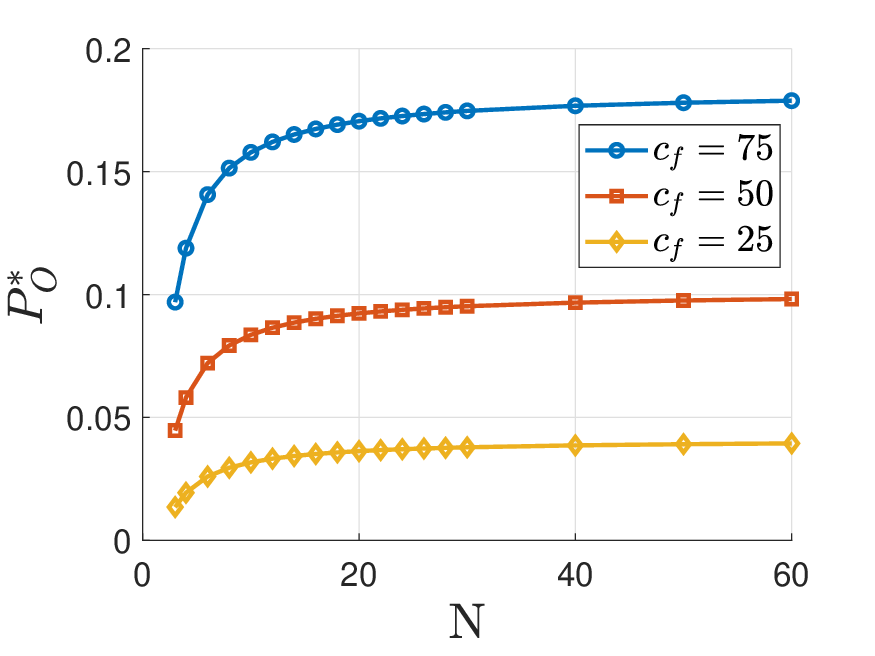}} 
   \hfill
    \subfloat[]{% 
        \includegraphics[width=0.3\linewidth]{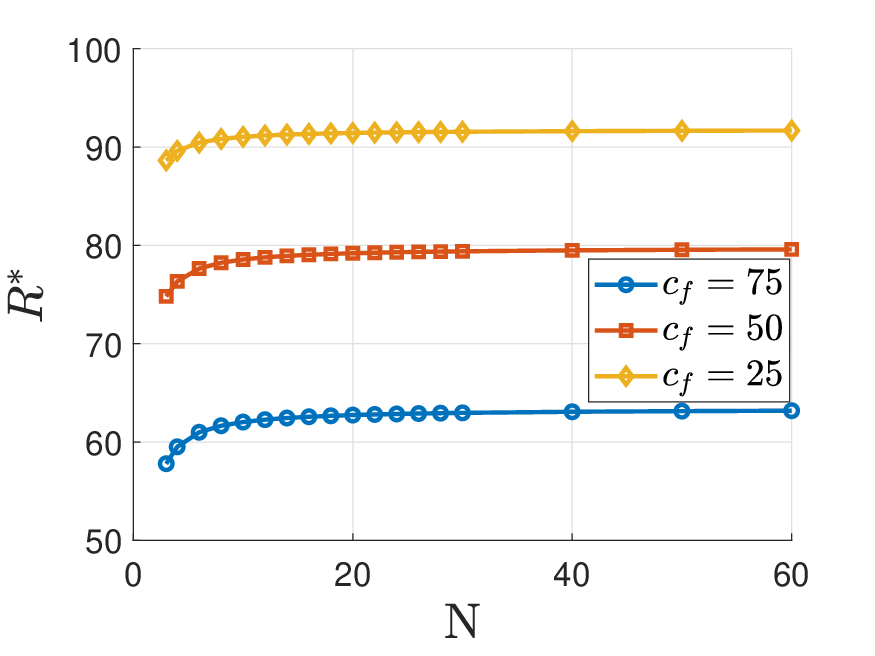}} 
    \hfill    
    \caption{Equilibrium upload probability $q_u^*$, outage probability $P_O^*$ and expected reward $R^*$ as a function of N, with $b=100$, $c_l=1$, $p=100$ and $c_f= 25,50$ and $75$.(a) Upload Probability at equilibrium. (b) Probability of outage at equilibrium. (c) Expected reward at equilibrium.} 
    \label{sysEval_wrt_N} 
\end{figure}

We begin by analyzing the equilibrium behavior of the system with respect to the number of players. We fix the reward at $b=100$, the penalty at $p=100$, and set $c_l=1$, while considering three values for the upload cost: $c_f=25, 50,$ and $75$. The results are presented in Figure~\ref{sysEval_wrt_N}.
\\
As shown in Figure~\ref{sysEval_wrt_N}.a, the equilibrium upload probability $q_u^*$ decreases as the number of relayers $N$ increases. This observation is consistent with the behavior of volunteer’s dilemma games, where the likelihood of an individual volunteering drops as the group size grows. Intuitively, the larger the set of participants, the more each individual relies on others to take action. Furthermore, as the upload cost $c_f$ increases, the equilibrium upload probability decreases, since higher costs discourage relayers from submitting the transaction to the blockchain.
Figure~\ref{sysEval_wrt_N}.b illustrates the corresponding outage probability at equilibrium $P_O^*$, which increases with both $N$ and $c_f$. Since the outage probability evolves in the opposite direction to the equilibrium upload probability, the results are consistent with expectations. A closer look shows that for smaller network sizes (e.g., $N \leq 15$), the system exhibits a clear trade-off between privacy and reliability. However, once the number of relayers exceeds a certain threshold, the outage probability stabilizes, with variations becoming very small (on the order of $0.001$) and continuing to decrease as $N$ grows. This indicates that additional relayers contribute to privacy, as they increase the anonymity set size, without further degrading reliability. Notably, when $c_f=25$, the system achieves the best reliability with an outage probability lower than $0.05$.
Finally, Figure~\ref{sysEval_wrt_N}.c presents the expected reward $R^*$ at equilibrium. As anticipated, lower upload costs substantially increase the expected rewards: with $c_f=25$, individual relayers obtain up to $91.65\%$ of the benefit $b$, compared to $79.49\%$ when $c_f=50$ and $63\%$ when $c_f=75$. Furthermore, $R^*$ shows only small variations as $N$ increases, indicating that an individual player would be indifferent to the overall size of the relayer set in terms expected payoffs.

\par
\begin{figure}[htbp] 
    \centering 
    \subfloat[]{% 
        \includegraphics[width=0.3\linewidth]{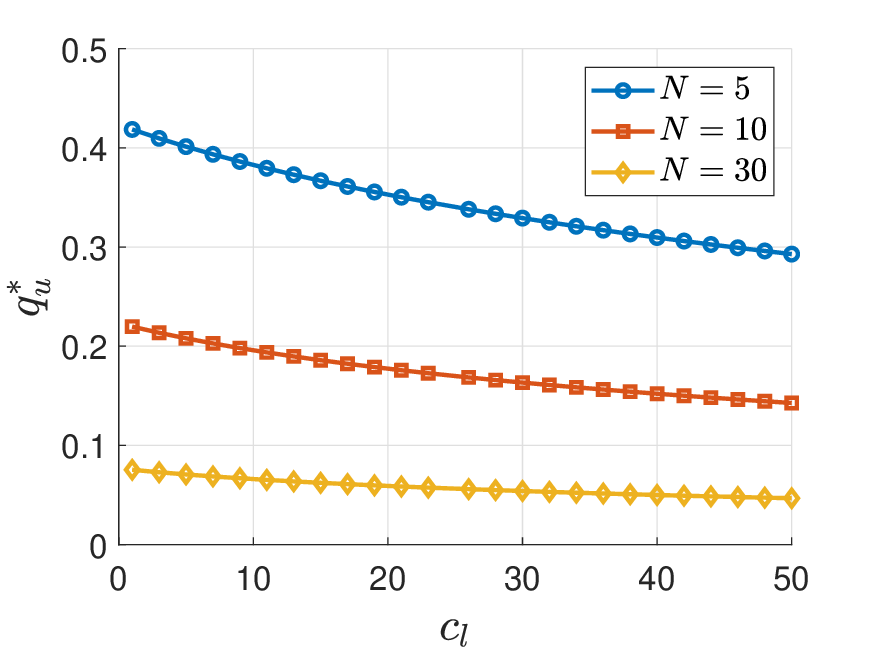}} 
        \hfill 
    \subfloat[]{% 
        \includegraphics[width=0.3\linewidth]{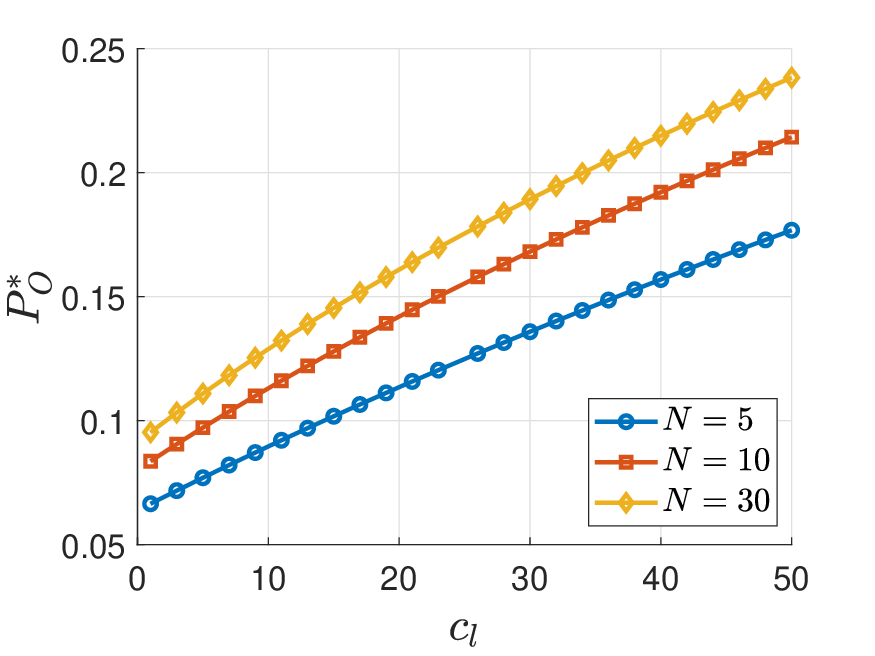}} 
        \hfill 
    \subfloat[]{% 
        \includegraphics[width=0.3\linewidth]{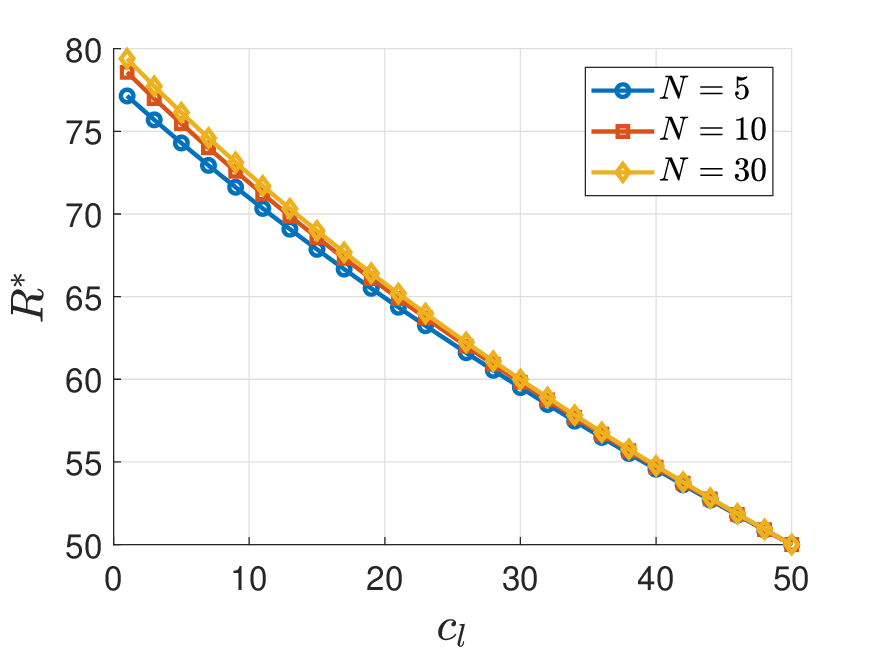}} 
        \hfill 
    \caption{Equilibrium upload probability $q_u^*$, outage probability $P_O^*$ and expected reward $R^*$ as a function of the unit of cost $c_l$, with $b=100$, $c_f=50$, $p=100$ and $N= 5,10$ and $30$.(a) Upload probability at equilibrium. (b) Probability of outage at equilibrium. (c) Expected reward at equilibrium.} 
    \label{sysEval_wrt_cl} 
\end{figure}

 
Next, we fix the number of relayers and the cost of the first upload $c_f$, and vary the cost of later uploads $c_l$ from a unit cost of $1$ up to $c_f=50$. The other parameters are set to $b=p=100$, and the metrics are evaluated for $N=5$, $N=10$, and $N=30$. Results are presented in Figure~\ref{sysEval_wrt_cl}. As shown in the figure, the higher the cost of reverted uploads, the lower the equilibrium upload probability. Indeed, higher $c_l$ discourages players from taking the initiative, reducing participation. Consequently, the reliability of the system degrades considerably as $c_l$ increases. For instance for $N=5$, when the upload cost equals half the intended benefit ($c_f=c_l=50$ and $b=100$), the outage probability rises from approximately $0.06$ to nearly $0.17$. Interestingly, in scenarios where the cost of interacting with the blockchain is high, the system’s reliability can still be improved by adjusting the penalty parameter $p$, as shown in the following results.
\par
\begin{figure}[htbp]
    \centering 
    \subfloat[]{% 
        \includegraphics[width=0.3\linewidth]{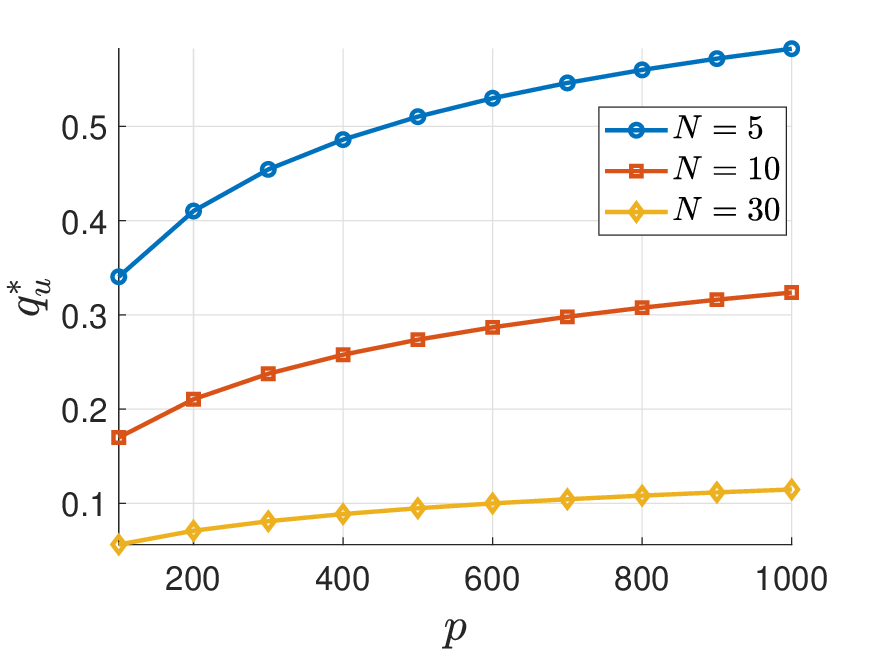}} 
        \hfill 
    \subfloat[]{% 
        \includegraphics[width=0.3\linewidth]{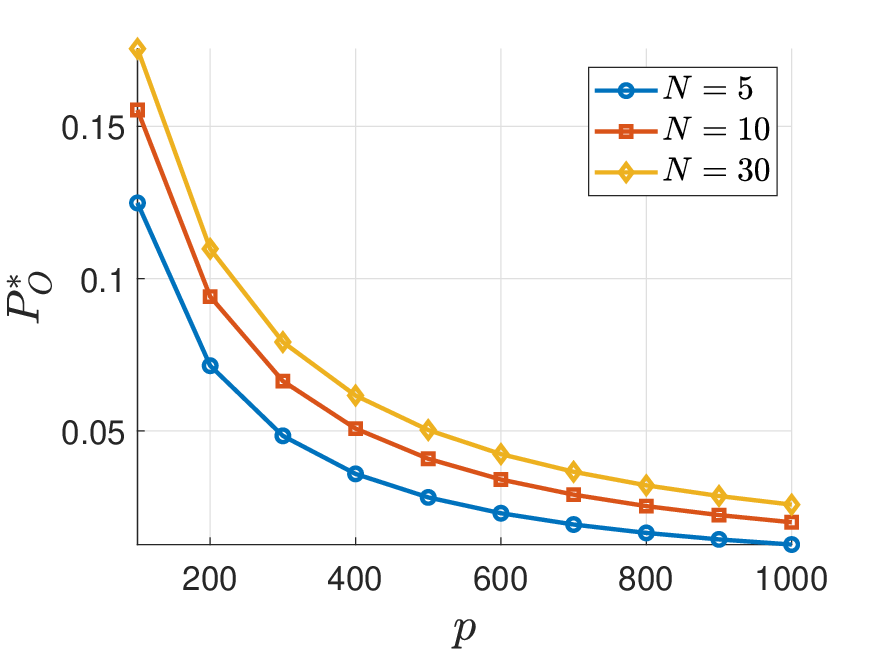}} 
        \hfill 
    \subfloat[]{% 
        \includegraphics[width=0.3\linewidth]{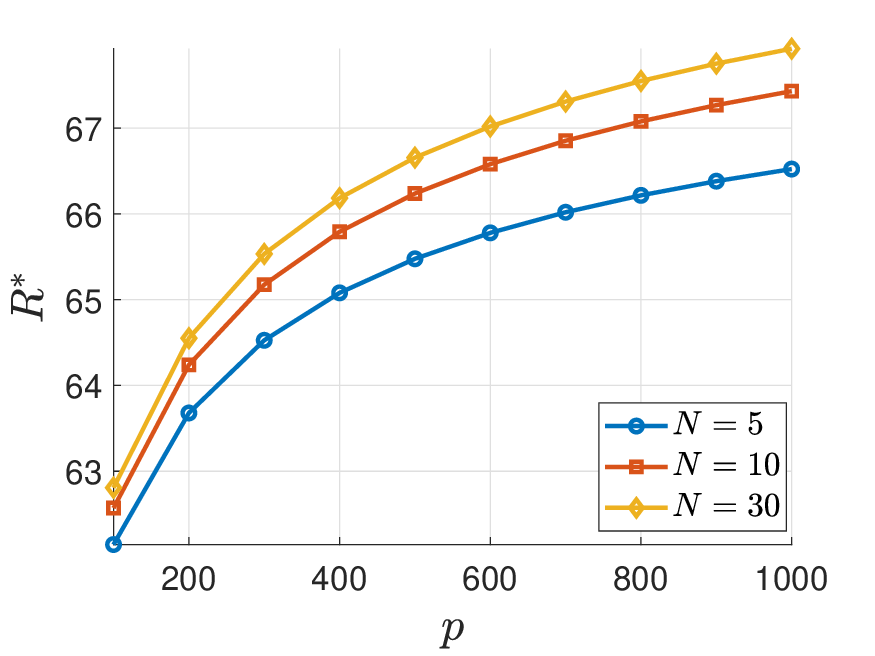}} 
        \hfill 
    \caption{Equilibrium upload probability $q_u^*$, outage probability $P_O^*$ and expected reward $R^*$ as a function of the penalty $p$, with $b=100$, $c_f=50$, $c_l=25$ and $N= 5,10$ and $30$.(a) Upload probability at equilibrium. (b) Probability of outage at equilibrium. (c) Expected reward at equilibrium.} 
    \label{sysEval_wrt_p} 
\end{figure}

We now analyze the impact of the penalty parameter $p$ on system performance. Specifically, we vary $p$ to evaluate how stronger penalties influence the equilibrium behavior when the cost of uploading is high, i.e., when the cost of the first upload is half the benefit and that of later uploads is a quarter: $b = 100$, $c_f = 50$, and $c_l = 25$. Results are shown in Figure~\ref{sysEval_wrt_p}. As illustrated in Figure~\ref{sysEval_wrt_p}.a, the equilibrium upload probability $q_u^*$ increases with $p$, as higher penalties for system outages strengthen the incentive for relayers to upload. Correspondingly, Figure~\ref{sysEval_wrt_p}.b shows that the outage probability $P_O^*$ decreases sharply as $p$ increases, confirming that penalty mechanisms effectively mitigate failures. For large penalty values, specifically, when $p$ exceeds five times the upload reward, the system achieves an outage probability lower than $0.05$ across all values of $N$ while maintaining a positive expected reward. Finally, Figure~\ref{sysEval_wrt_p}.c shows that the expected equilibrium reward $R^*$ increases steadily with $p$, as stronger penalties discourage free-riding behavior and enhance overall system reliability.

\begin{figure}[htbp]
    \centering 
    \subfloat[]{% 
        \includegraphics[width=0.3\linewidth]{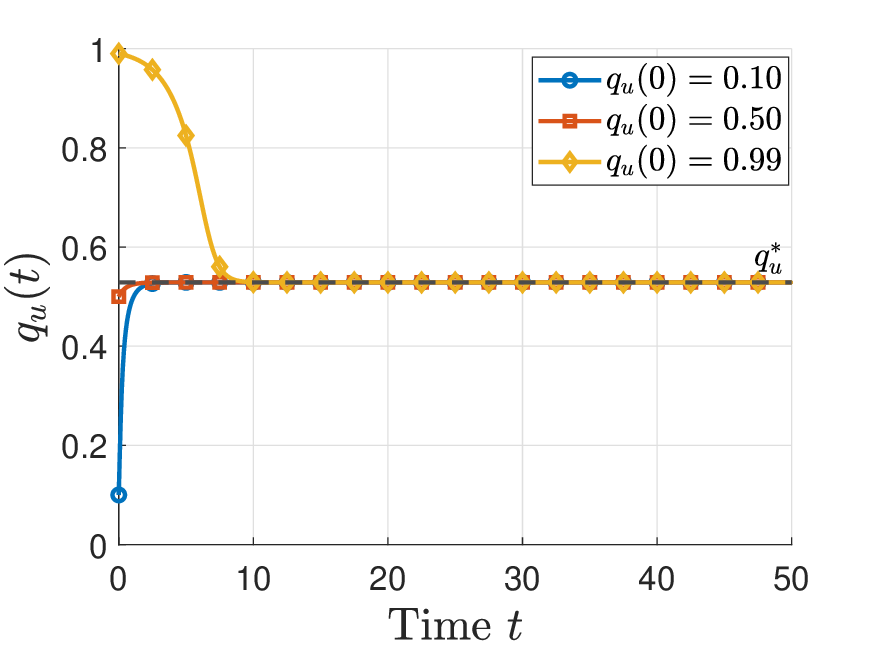}} 
        \hfill 
    \subfloat[]{% 
        \includegraphics[width=0.3\linewidth]{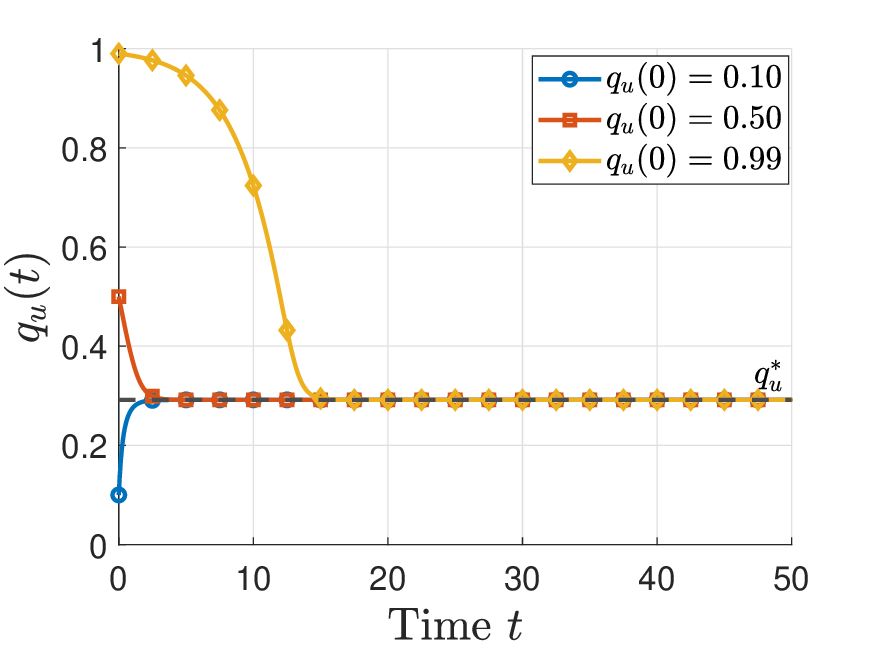}} 
        \hfill 
    \subfloat[]{% 
        \includegraphics[width=0.3\linewidth]{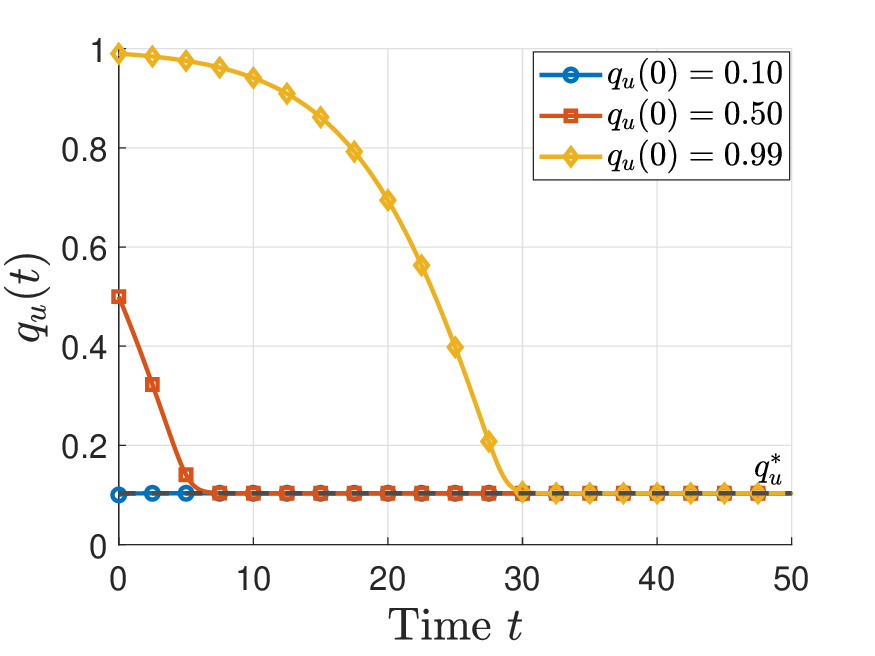}} 
        \hfill 
    \caption{Evolution of the upload probability $q_u(t)$ under the replicator’s dynamic with parameters $b=100$, $c_f=25$, $c_l=1$, $p=100$, and $\mu=0.1$. The trajectories are shown for different initial states $q_u(0)= 0.10$,$0.50$ and for different values of $N$. (a) $N=5$, (b) $N=10$, and (c) $N=30$.}
    \label{replicator_dynamics} 
\end{figure}

Next, we numerically evaluate the stability of the system. We begin by studying how it reacts to perturbations to the upload probability $q_u(t)$. To do this, we simulate the replicator dynamic under different initial conditions and numbers of relayers $N$. The parameters are set to $b = 100$, $c_f = 25$, $c_l = 1$, $p = 100$, and $\mu = 0.1$. Results are shown in Figure~\ref{replicator_dynamics}. Across all cases, the trajectories converge to the unique symmetric Nash equilibrium $q_u^*$, confirming that the equilibrium is globally asymptotically stable, as stated in Proposition~\ref{stability}. However, the convergence rate varies depending on both the initial state and the number of players: when the initial probability is far from equilibrium, the system requires more time to stabilize, and larger populations $N$ lead to slower convergence.
\par
 
\begin{figure}[htbp]
    \centering 
    \subfloat[]{% 
        \includegraphics[width=0.3\linewidth]{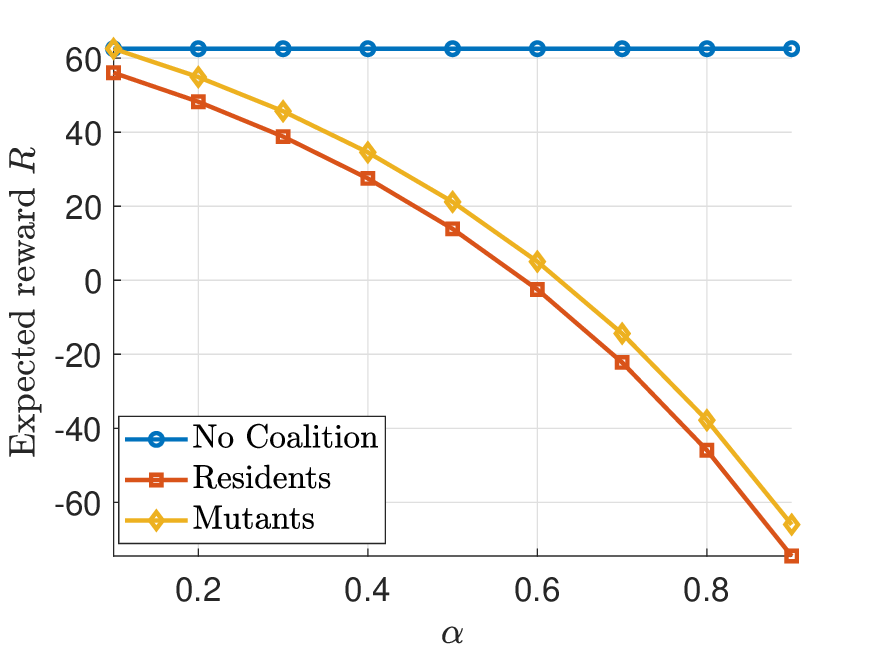}} 
        \hfill 
    \subfloat[]{% 
        \includegraphics[width=0.3\linewidth]{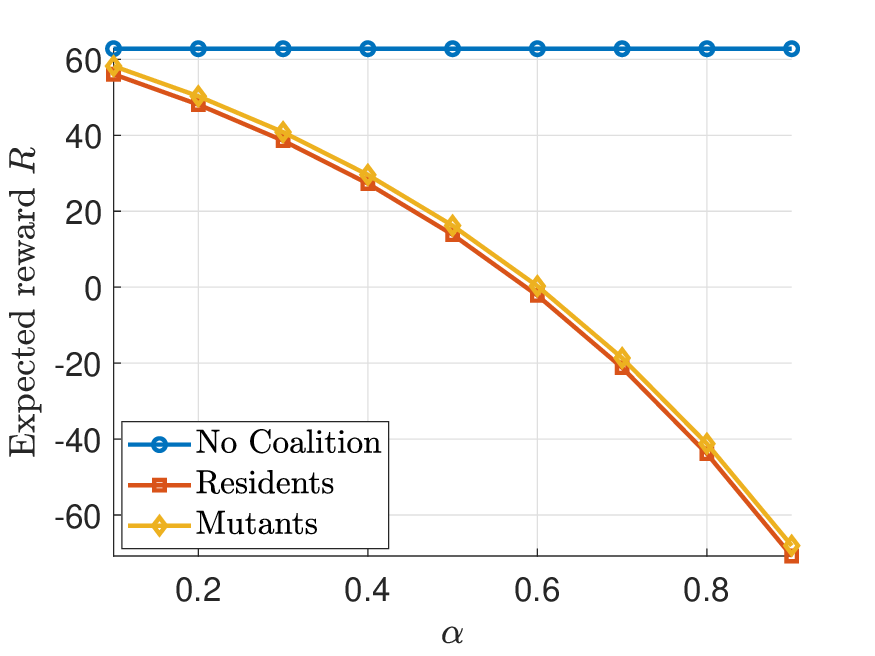}} 
        \hfill 
    \subfloat[]{% 
        \includegraphics[width=0.3\linewidth]{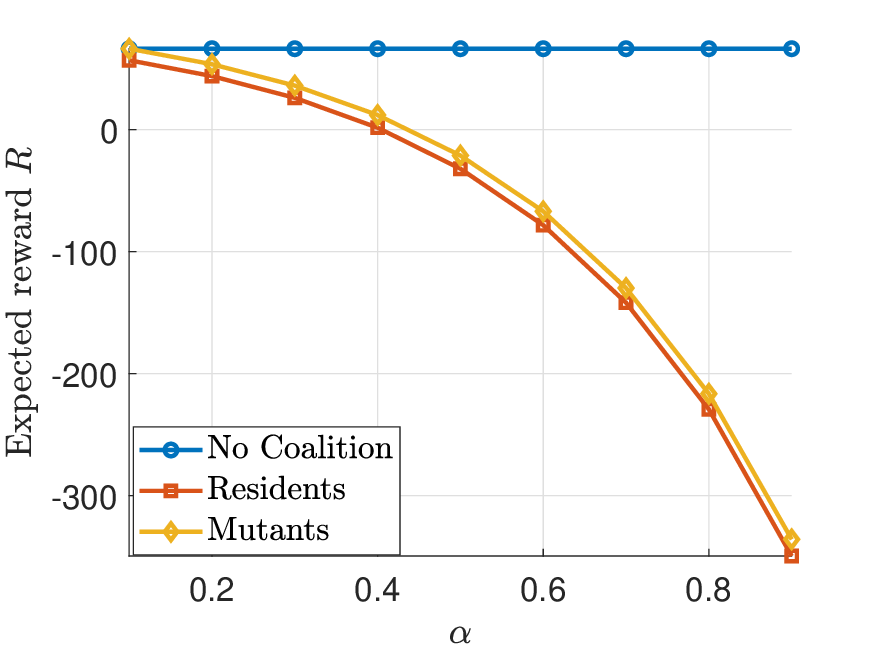}} 
        \hfill 
     \subfloat[]{% 
        \includegraphics[width=0.3\linewidth]{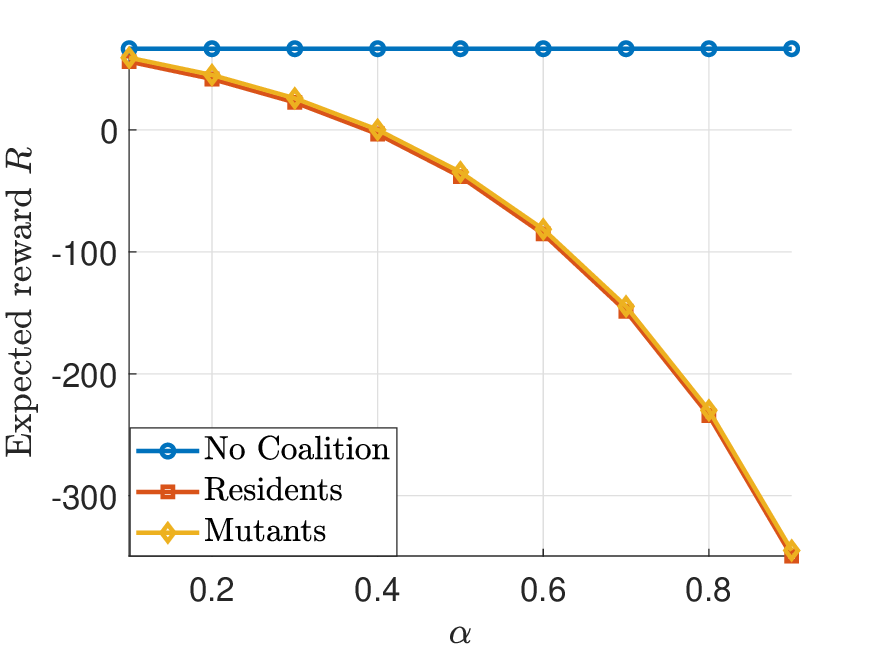}} 
        \hfill 
    \caption{Comparison of player rewards under two scenarios: (i) all players follow the equilibrium strategy, and (ii) a coalition of players deviates while the remaining players adhere to equilibrium, with $b = 100$, $c_f = 50$, $c_l = 25$. (a) $N=10$, $p=100$ (b) $N=30$, $p=100$ (c) $N=10$, $p=500$ and (d) $N=30$, $p=500$.}
    \label{strong_equilibrium} 
\end{figure}

Figure~\ref{strong_equilibrium} analyzes coalition deviations in which a fraction $\alpha \in (0,1)$ of the players abandon the equilibrium strategy and instead commit to the pure action of not uploading, while the remaining players, referred to as residents, continue to play the equilibrium strategy. The baseline, shown as the flat blue curves, corresponds to the scenario with no coalition, where all players follow the equilibrium strategy. Results show that no coalition size, where mutants never upload, is able to increase its expected reward.  Moreover, the threshold of mutants beyond which the expected payoff becomes negative depends on the penalty parameter. In this numerical evaluation, we consider a high-cost scenario with $c_f = 50$ and $c_l = 25$. When the penalty is set equal to the benefit ($p = 100$), both mutants and residents start losing their stakes at approximately $\alpha \approx 0.6$, while for a stronger penalty ($p = 500$), the threshold decreases to around $\alpha \approx 0.4$. The number of players has little impact on this threshold, but it does affect the gap between the expected rewards of the mutant group and the resident group. When $N = 30$, both mutants and residents achieve almost the same reward. This threshold is interesting when we consider unintentional failures, rather than strategic deviations, as it quantifies the robustness of the system against failures. 

\subsection{Discussion}

The theoretical analysis and numerical evaluations confirm that the proposed incentive mechanism for decentralized relayers admits a stable and robust equilibrium, resilient to both perturbations and coordinated deviations. At the same time, the equilibrium behavior induces inherent trade-offs between privacy, reliability, robustness, and cost. We discuss these trade-offs and their practical implications below.

\begin{enumerate}
    \item \textit{Resilience to deviations:} As the system parameters are fixed and publicly known in advance, the participating relayers can be configured to upload with the recommended probability $q_u^*$, which represents the system’s equilibrium. In this case, no rational individual can increase their utility by deviating unilaterally (Nash equilibrium), nor can any coalition of players improve their collective outcome by deviating jointly ($\alpha-$ strong Nash equilibrium). Furthermore, when the system experiences perturbations, if nodes are configured to adjust their upload probabilities through an evolutionary dynamic, they will eventually converge back to the equilibrium, ensuring the system’s resilience to deviations. Finally, even when deviations occur unintentionally due to node failures rather than strategic manipulation, the results show that players continue to earn positive rewards up to threshold $\alpha$ of failing nodes, demonstrating the system’s robustness.
    \item \textit{The privacy-reliability-robustness tradeoff:} In the proposed system, increasing the number of relayers enlarges the anonymity set, thereby strengthening privacy by making it more difficult to infer the origin of a transaction. Nevertheless, this improvement in privacy comes at the expense of reliability, as the probability of outage increases with the number of participating relayers. To mitigate this effect, the penalty parameter can be tuned to incentivize uploads and maintain system performance. Numerical evaluations confirm that, for sufficiently high penalty values, the system sustains a low outage probability even when the number of relayers and upload costs are high. On the other hand, increasing the penalty reduces the robustness of the system. Indeed, when penalties are high, the threshold of failing nodes from which the relayer's expected payoff becomes negative decreases. Hence, the system exhibits an inherent trade-off between privacy, reliability, and robustness, where improving one dimension inevitably affects the others.
    \item \textit{The privacy–cost tradeoff:} Numerical results show that beyond a certain number of relayers, increasing the size of the relayer set has only a marginal impact on the outage probability. This observation suggests that the anonymity set can be expanded to enhance privacy without significantly compromising reliability or robustness. Nevertheless, it is important to note that a larger relayer set increases the overall operational cost of the system. Since each node receives a reward equal to the benefit $b$, the total cost associated with rewarding the relayers becomes $N \times b$. Therefore, although enlarging the anonymity set effectively enhances privacy and decentralization, it also imposes additional financial overhead. This implies that the relayer set cannot be expanded indefinitely without compromising the system’s economic feasibility.
    \item \textit{Model limitations and deployment challenges:}  
    The proposed framework abstracts several practical aspects of decentralized relayer systems. First, the privacy analysis focuses on the blockchain layer and assumes an observer limited to on-chain information; network-level adversaries capable of traffic analysis, timing correlation, or message interception are not modeled. Addressing such threats would require combining the proposed incentive mechanism with network-layer anonymity techniques, such as controlled gossiping or mixnets. Second, upload costs are modeled as normalized utilities rather than exact gas measurements. While this captures the qualitative distinction between successful and reverted uploads, the precise ratio between $c_f$ and $c_l$ depends on contract design and execution paths. Finally, the analysis focuses on a one-shot interaction with static parameters. In real deployments, relayers interact repeatedly under fluctuating gas prices and dynamic participation. While the evolutionary stability results suggest convergence toward equilibrium over time, extending the model to explicitly capture repeated interactions with behavioral strategies and time-varying costs remains an important direction for future work.
\end{enumerate}

\section{Conclusion}
In this work, we proposed a game-theoretic incentive mechanism for decentralized blockchain relayers, designed to enhance privacy while maintaining system reliability. By modeling the interactions among relayers as a non-cooperative game inspired by the Volunteer's Dilemma, we established the existence, uniqueness, and evolutionary stability of a mixed-strategy Nash equilibrium. Numerical evaluations demonstrated that increasing the number of relayers improves privacy by enlarging the anonymity set but decreases reliability as the outage probability rises. Moreover, stronger penalty mechanisms effectively reduce coordination failures, though at the expense of the overall robustness. The proposed framework provides a practical design paradigm for privacy-preserving and incentive-compatible decentralized relayer architectures in blockchain systems. The results reveal a fundamental trade-off between privacy, reliability, robustness, and cost. Future work will  extend the model to consider scenarios in which the number of relayers is stochastic rather than fixed and to include not only blockchain uploads but also the incentives for message dissemination among relayers.

\bibliographystyle{elsarticle-num}
\bibliography{references}

@article{volunteersDilemma85,
author = {Diekmann, Andreas},
year = {1985},
month = {12},
pages = {605-610},
title = {Volunteer's Dilemma},
volume = {29},
journal = {Journal of Conflict Resolution - J CONFLICT RESOLUT},
doi = {10.1177/0022002785029004003}
}

@article{HOFBAUER2022,
author = {Hofbauer, Josef and Sorger, Gerhard},
year = {2011},
month = {11},
pages = {},
title = {A differential game approach to evolutionary equilibrium selection},
volume = {04},
journal = {International Game Theory Review},
doi = {10.1142/S0219198902000525}
}

@incollection{SANDHOLM,
title = {Chapter 8 - Local Stability under Evolutionary Dynamics},
booktitle = {Population games and evolutionary dynamics},
publisher = {The MIT press},
pages = {271-317},
year = {2011},
isbn = {978-0-262-19587-4},
author = {Sandholm, William H},
}

@incollection{SAMSON_CHAP1,
title = {Chapter 1 - A Very Short Tour of Game Theory },
author = {Samson Lasaulce and Hamidou Tembine},
booktitle = {Game Theory and Learning for Wireless Networks},
publisher = {Academic Press},
address = {Oxford},
pages = {3-40},
year = {2011},
isbn = {978-0-12-384698-3},
doi = {https://doi.org/10.1016/B978-0-12-384698-3.00003-7},

}

@incollection{SAMSON_CHAP2,
title = {Chapter 2 - Playing with Equilibria in Wireless Non-Cooperative Games },
author = {Samson Lasaulce and Hamidou Tembine},
booktitle = {Game Theory and Learning for Wireless Networks},
publisher = {Academic Press},
address = {Oxford},
pages = {41-68},
year = {2011},
isbn = {978-0-12-384698-3},
doi = {https://doi.org/10.1016/B978-0-12-384698-3.00003-7},

}

@misc{relayers_def,
    title = {Relayers – OpenZeppelin Documentation},
    author = {{OpenZeppelin}},
    year = {2025},
    note = {\url{https://docs.openzeppelin.com/defender/module/relayers}, Accessed: 2025-09-25},
}

@ARTICLE{survey_privacy,
  author={Wang, Dan and Zhao, Jindong and Wang, Yingjie},
  journal={IEEE Access}, 
  title={A Survey on Privacy Protection of Blockchain: The Technology and Application}, 
  year={2020},
  volume={8},
  number={},
  pages={108766-108781},
  doi={10.1109/ACCESS.2020.2994294}}

@INPROCEEDINGS{deanonymization,
  author={Gao, Yue and Shi, Jinqiao and Wang, Xuebin and Shi, Ruisheng and Yin, Zelin and Yang, Yanyan},
  booktitle={2021 IEEE Intl Conf on Parallel \& Distributed Processing with Applications, Big Data \& Cloud Computing, Sustainable Computing \& Communications, Social Computing \& Networking (ISPA/BDCloud/SocialCom/SustainCom)}, 
  title={Practical Deanonymization Attack in Ethereum Based on P2P Network Analysis}, 
  year={2021},
  volume={},
  number={},
  pages={1402-1409},
  doi={10.1109/ISPA-BDCloud-SocialCom-SustainCom52081.2021.00191}}

@inproceedings{flexible_privacy,
  title={A flexible network approach to privacy of blockchain transactions},
  author={M{\"o}dinger, David and Kopp, Henning and Kargl, Frank and Hauck, Franz J},
  booktitle={2018 IEEE 38th International Conference on Distributed Computing Systems (ICDCS)},
  pages={1486--1491},
  year={2018},
  organization={IEEE}
}

@article{active_passive,
  title={Toward active and passive confidentiality attacks on cryptocurrency off-chain networks},
  author={Nisslmueller, Utz and Foerster, Klaus-Tycho and Schmid, Stefan and Decker, Christian},
  journal={arXiv preprint arXiv:2003.00003},
  year={2020}
}

@INPROCEEDINGS{zerocash2014,
  author={Ben Sasson, Eli and Chiesa, Alessandro and Garman, Christina and Green, Matthew and Miers, Ian and Tromer, Eran and Virza, Madars},
  booktitle={2014 IEEE Symposium on Security and Privacy}, 
  title={Zerocash: Decentralized Anonymous Payments from Bitcoin}, 
  year={2014},
  volume={},
  number={},
  pages={459-474},
  doi={10.1109/SP.2014.36}}

@article{dandelionpp,
author = {Fanti, Giulia and Venkatakrishnan, Shaileshh Bojja and Bakshi, Surya and Denby, Bradley and Bhargava, Shruti and Miller, Andrew and Viswanath, Pramod},
title = {Dandelion++: Lightweight Cryptocurrency Networking with Formal Anonymity Guarantees},
year = {2018},
issue_date = {June 2018},
publisher = {Association for Computing Machinery},
address = {New York, NY, USA},
volume = {2},
number = {2},
doi = {10.1145/3224424},
journal = {Proc. ACM Meas. Anal. Comput. Syst.},
month = jun,
articleno = {29},
numpages = {35},
}

@inproceedings{loopix,
author = {Piotrowska, Ania M. and Hayes, Jamie and Elahi, Tariq and Meiser, Sebastian and Danezis, George},
title = {The loopix anonymity system},
year = {2017},
isbn = {9781931971409},
publisher = {USENIX Association},
address = {USA},
booktitle = {Proceedings of the 26th USENIX Conference on Security Symposium},
pages = {1199–1216},
numpages = {18},
location = {Vancouver, BC, Canada},
series = {SEC'17}
}

@misc{zklogin,
  author       = "{Sui Documentation}",
  title        = "{zkLogin - Security and Privacy}",
  howpublished = "\url{https://docs.sui.io/concepts/cryptography/zklogin#security-and-privacy}",
  note         = "Accessed: 2025-07-28"
}

@INPROCEEDINGS{nymCredentials,
  author={Halpin, Harry},
  booktitle={2020 Crypto Valley Conference on Blockchain Technology (CVCBT)}, 
  title={Nym Credentials: Privacy-Preserving Decentralized Identity with Blockchains}, 
  year={2020},
  volume={},
  number={},
  pages={56-67},
  doi={10.1109/CVCBT50464.2020.00010}}

@article{guerraouiPrivacy,
  title={On the inherent anonymity of gossiping},
  author={Guerraoui, Rachid and Kermarrec, Anne-Marie and Kucherenko, Anastasiia and Pinot, Rafael and Voitovych, Sasha},
  journal={arXiv preprint arXiv:2308.02477},
  year={2023}
}

@misc{nymNetwork,
  author       = "{Claudia D{\'i}az and Harry Halpin and Aggelos Kiayias}",
  title        = "{The Nym Network The Next Generation of Privacy Infrastructure}",
  year={2021},
  howpublished = "\url{https://api.semanticscholar.org/CorpusID:233218535}",
  note         = "Accessed: 2025-07-28"
}

@INPROCEEDINGS{zkRelay,
  author={Westerkamp, Martin and Eberhardt, Jacob},
  booktitle={2020 IEEE European Symposium on Security and Privacy Workshops (EuroS\&PW)}, 
  title={zkRelay: Facilitating Sidechains using zkSNARK-based Chain-Relays}, 
  year={2020},
  volume={},
  number={},
  pages={378-386},
  doi={10.1109/EuroSPW51379.2020.00058}}

@misc{openzeppelin_relayers,
  author       = {{OpenZeppelin}},
  title        = {Defender Relayers Documentation},
  howpublished = {\url{https://docs.openzeppelin.com/defender/module/relayers}},
  note         = "Accessed: 2025-07-28"
}

@misc{EIP4337,
  author       = {Vitalik Buterin and Yoav Weiss and Dror Tirosh and Shahaf Nacson and Alex Forshtat and Kristof Gazso and Tjaden Hess},
  title        = {{ERC-4337: Account Abstraction Using Alt Mempool [DRAFT]}},
  howpublished = {\emph{Ethereum Improvement Proposals}, no. 4337},
  year         = {2021},
  month        = {September},
  note         = {[Online]. Available: \url{https://eips.ethereum.org/EIPS/eip-4337}}
}

@misc{tornado,
  author       = {{Tornado Cash}},
  title        = {Tornado.Cash Documentation},
  year         = {2024},
  url          = {https://docs.tornado.ws/},
  note         = {Accessed: 2025-08-22}
}

@article{replicatorSymmetric,
  title={The replicator equation and other game dynamics},
  author={Cressman, Ross and Tao, Yi},
  journal={Proceedings of the National Academy of Sciences},
  volume={111},
  number={supplement\_3},
  pages={10810--10817},
  year={2014},
  publisher={National Academy of Sciences}
}

@article{evolutionaryGameDynamic,
  title={Evolutionary game dynamics},
  author={Hofbauer, Josef and Sigmund, Karl},
  journal={Bulletin of the American mathematical society},
  volume={40},
  number={4},
  pages={479--519},
  year={2003}
}

@Article{goncalves2022dandelionSecurity,
AUTHOR = {Goncalves, Brian and Mashatan, Atefeh},
TITLE = {On the Security of the Dandelion Protocol},
JOURNAL = {Mathematics},
VOLUME = {10},
YEAR = {2022},
NUMBER = {7},
ARTICLE-NUMBER = {1054},
ISSN = {2227-7390},
DOI = {10.3390/math10071054}
}

@ARTICLE{vd_vanet,
  author={Limouchi, Elnaz and Mahgoub, Imad},
  journal={IEEE Transactions on Intelligent Transportation Systems}, 
  title={Volunteers Dilemma Game Inspired Broadcast Scheme for Vehicular Ad Hoc Networks}, 
  year={2019},
  volume={20},
  number={12},
  pages={4439-4449},
  doi={10.1109/TITS.2018.2886232}}

@article{vd_naja,
title = {Adaptive alert content dissemination protocol inspired from volunteer's dilemma game for Vehicular Ad-hoc Networks},
journal = {Vehicular Communications},
volume = {19},
pages = {100180},
year = {2019},
issn = {2214-2096},
doi = {https://doi.org/10.1016/j.vehcom.2019.100180},
author = {Assia Naja and Mohammed Boulmalf and Mohamed Essaaidi and Omar {Ait Oualhaj}},
}

@article{myatt_oxford_2008,
title = {An evolutionary analysis of the volunteer's dilemma},
journal = {Games and Economic Behavior},
volume = {62},
number = {1},
pages = {67-76},
year = {2008},
issn = {0899-8256},
doi = {https://doi.org/10.1016/j.geb.2007.03.005},
author = {David P. Myatt and Chris Wallace},
}

@article{konrad,
  title={The volunteer’s dilemma in finite populations},
  author={Konrad, Kai A and Morath, Florian},
  journal={Journal of Evolutionary Economics},
  volume={31},
  number={4},
  pages={1277--1290},
  year={2021},
  publisher={Springer}, 
  doi = {https://doi.org/10.1007/s00191-020-00719-y}
}

\appendices  
\section{ Appendix A - Proofs}  
\subsection{Proof of Theorem \ref{uniquenessNE}}
\begin{proof} 
Consider the relayer upload game $\mathcal{G}_N$.

\textbf{Non-existence of symmetric pure Nash equilibria.}
When all players choose not to upload, then the best response for any individual player is to deviate and upload, changing their utility from $-p<0$ to $b - c_f>0$. Similarly,  if all players choose to upload, then the best response is to not upload, and thus receiving a reward of $b$, which is preferable to either $b - c_f$ or $b - c_l$. Therefore, in the symmetric case, no pure strategy equilibrium exists. 

\par 
\textbf{Existence of a symmetric mixed Nash equilibrium.}
 We first compute the expected utility for a player $i \in \mathcal{N}$ when choosing to upload versus not upload and prove that they are equal at exactly one point in the open interval $(0,1)$.
\\
Assuming that all players upload with the same probability $0<q_u$, the expected utility of player $i$, taking action U, is given by :
\begin{equation}
    v(\textrm{U}, q_u) = q_f(b - c_f) + (1 - q_f)(b - c_l),
    \label{expectedU}
\end{equation}
where $q_f$ is the probability that the player $i$ is selected first, given that they chose to upload.
\par
Let $M$ be the random variable representing the total number of nodes which attempt to upload. $M$ follows a binomial distribution of parameters $N$ and $q_u$. We assume that the first one is selected  uniformly at random among the uploading nodes. The probability that a given node is selected first, given  that it chose to upload and that there are $m$ other nodes also attempting to upload, can be expressed as:
\begin{equation}
    \begin{aligned}
        q_f &= \sum\limits_{m=0}^{N-1} \frac{1}{m + 1}\binom{N-1}{m} q_u^m (1 - q_u)^{N - 1 - m} \\
            &= \frac{1}{N q_u}(1 - (1 - q_u)^N).
    \end{aligned}
\label{qf}
\end{equation}
\par
On the other hand, the expected utility of playing NU is given by : 
\begin{equation}
    v(\textrm{NU},q_u) = b(1-(1-q_u)^{{N - 1}}) - p(1-q_u)^{N - 1}.
    \label{expectedNU}
\end{equation}

\par
To identify a symmetric mixed strategy Nash equilibrium, we apply the indifference principle.
\\ Given that all other players upload with probability  $q_u$, this leads to the condition :
$
v(\text{U}, q_u) = v(\text{NU}, q_u),
$  
which is equivalent to finding a root of the function  
$
g(q_u) = v(\text{U}, q_u) - v(\text{NU}, q_u),
$  
where $g(q_u^*) = 0$  for some $q_u^* \in (0,1)$. A symmetric mixed equilibrium thus exists if and only if $g$ has a root in the open interval $(0,1)$.

\par 
We examine the signs of $g$ at the extreme points of the interval. At $q_u = 0$, where all other players choose not to upload, we have:
\begin{equation*}
    \begin{aligned}
        g(0) &= v(\text{U},\text{NU},\text{NU},...,\text{NU}) - v(\text{NU},\text{NU},\text{NU},...,\text{NU}) \\
            &= b - c_f + p > 0.
    \end{aligned}
\end{equation*}
At $q_u = 1$, where all players upload, we obtain:
\begin{equation*}
    \begin{aligned}
        g(1) &=v(\text{U},\text{U},\text{U},...,\text{U}) -  v(\text{NU},\text{U},\text{U},...,\text{U}) \\
             &= -c_l - \frac{1}{N}(c_f-c_l) < 0. 
    \end{aligned}
\end{equation*}
By continuity of \( g \), there exists at least one root \( q_u^* \in (0,1) \). Hence, a symmetric mixed strategy Nash equilibrium exists.

\par
\textbf{Uniqueness of the symmetric mixed Nash equilibrium.}  
 We consider a simplified expression of $g$, denoted $h$, as defined in Theorem~\ref{uniquenessNE}. This function is obtained by substituting the explicit forms of the utilities from equations~\ref{expectedU} and~\ref{expectedNU} into the definition of $g$. Uniqueness then follows by showing that $h'$ changes sign exactly once in the interval $(0,1)$.
\par 
Using Equations~\ref{expectedU}, \ref{qf}, and \ref{expectedNU}:
\begin{equation*}
    \begin{split}
        h(q_u)  &= Nq_u(b+p)(1-q_u)^{N-1} - N q_u c_l \\
                & + (c_l - c_f) (1-(1-q_u)^{N} ),
    \end{split} 
\end{equation*}
The derivative of $h$ is:
\begin{equation}
\begin{aligned}
h^{'}(q_u) &= N(b+p)\left[(1-q_u)^{N-1} - q_u(N-1)(1-q_u)^{N-2}\right] \\
           &\quad - Nc_l + N(c_l - c_f)(1 - q_u)^{N-1}, 
\end{aligned}
\label{first_derivative}
\end{equation}

and the second derivative is:
\begin{equation}
    \begin{aligned}
        h^{''}(q_u) &= N(N-1)(b + p)(1 - q_u)^{N-3} [N q_u \\
                    &\quad - 2 - \frac{c_l - c_f}{b + p}(1 - q_u)].
    \end{aligned} 
\label{second_derivative}
\end{equation}

Let $\hat{q}_u$ denote the root of $h^{''}(q_u) = 0$, given by:

\begin{equation*}
        \hat{q}_u = \frac{2(b + p) + (c_l - c_f)}{N(b + p) + (c_l - c_f)}.
\end{equation*}

Since $N \geq 3$ and $b>c_f$, $\hat{q}_u$ is well-defined.

\par 

We now study the sign of $h'$ over the intervals $(0, \hat{q}_u)$ and $(\hat{q}_u, 1)$:

\begin{itemize}
   
    \item On $(0, \hat{q}_u)$, we have $h''(q_u) < 0$, since the term 
    $
    Nq_u - 2 - \frac{c_l - c_f}{b + p}(1 - q_u) < 0
    $, 
    for all $q_u < \hat{q}_u$. Thus, $h' $ is strictly decreasing. Given that $h'(0) > 0 $ and $ h'(\hat{q}_u) < 0 $, there exists a unique solution of  $h'(q_u) = 0$ in $(0, \hat{q}_u)$ .
    \\
    \item On $(\hat{q}_u, 1)$,  $h''(q_u) > 0$, and $h'$ is strictly increasing. Since $ h'(\hat{q}_u) < 0 $ and $h'(1) = -Nc_l < 0 $, we have $ h'(q_u) < 0 $ throughout this interval. Thus, no other root of $ h' $ exists.
    \\
\end{itemize}
Therefore, $h'$ changes sign exactly once in the open interval $(0,1)$. Hence, $h$ crosses zero exactly once in the open interval $(0,1)$.

\par 
It follows that the game $\mathcal{G}_N$ admits a unique mixed NE in the open interval $(0,1)$. 
\end{proof}
\subsection{Proof of Proposition \ref{stability} }
\begin{proof}

We first show that the game $\mathcal{G}_N$ is a potential game.

To begin, we note that $\mathcal{G}_N$ is a symmetric game. Indeed, all players share the same action space, and the payoff functions are invariant under any permutation of players' indices and actions. Furthermore, since each player has exactly two strategies, it is known that any symmetric $N$-player game with a common two-strategy set is an exact potential game \cite[Section~4.2.2]{HOFBAUER2022}. 

In particular, the function
\[
\phi(q_u) = \int_0^{q_u} g(x)\,dx + C
\]
 is a valid potential function for $\mathcal{G}_N$, where $g$ is defined in Equation~\ref{defg} and $C \in \mathbb{R}$ is a constant. Indeed, the derivative of $\phi$ is g. Hence, maximizing the potential function is actually solving the root of $g$. 

Moreover, since $\mathcal{G}_N$ is a potential game, the asymptotic stability of the Nash equilibrium can be established by invoking a standard result from evolutionary game theory: any isolated local maximizer of a potential function is asymptotically stable under any evolutionary dynamic that satisfies positive correlation and Nash stationarity \cite[Theorem~8.2.1]{SANDHOLM}. 
\par
Therefore, the symmetric mixed-strategy Nash equilibrium $q_u^*$ is locally asymptotically stable under any such dynamic. 
\end{proof}
\subsection{Proof of Proposition \ref{propess1}}
\begin{proof}
Let $\underline{a} \in \mathrm{SE}$ be a profile with exactly one uploader, say player $i$. Observe that $b$ is the maximum possible payoff any player can achieve, which will be useful in evaluating potential deviations.

\textbf{Single-player deviations.}  
We first consider deviations by individual players. If the uploader $i$ were to switch from uploading to not uploading, then no player would be uploading in the resulting profile. This would yield a payoff of $-p$ for the uploader, which is strictly less than the original payoff $b - c_f$. Hence, the uploader has no incentive to deviate unilaterally. Similarly, if any non-uploader $j \neq i$ decides to upload, that player becomes a non-first uploader, receiving a payoff of $b - c_f$ or $b - c_l$, which is strictly less than $b$, the payoff they originally had. Therefore, no single player can improve their payoff by deviating alone.

\textbf{Coalitional deviations.}  
Next, we consider deviations by groups of players. Any coalition that includes the uploader cannot strictly improve all members’ payoffs, because if the uploader switches to not uploading, at least one member of the coalition would receive $-p$ (actually all of them will get $-p$ in that case), which is strictly worse than their original payoff of $b-c_f $ (or $b$). Likewise, any coalition composed entirely of non-uploaders cannot strictly improve all members simultaneously. If any non-uploader in the coalition switches to uploading, that player would become a non-first uploader and receive $b - c_l$, which is strictly less than $b$. Consequently, no coalition can strictly improve the payoffs of all its members by deviating.

\textbf{Exclusion of other profiles.}  
We consider profiles outside of the one-uploader structure. If no players upload, then the payoff for each player is $-p$, and a singleton coalition consisting of any player can improve their payoff by choosing to upload, achieving $b - c_f > -p$. Hence, profiles with zero uploaders are not strong equilibria. On the other hand, if multiple players upload, then the non-first uploaders receive $b - c_l < b$, and a coalition of these non-first uploaders can improve their payoffs by jointly switching to not uploading. Therefore, profiles with multiple uploaders are also not strong equilibria. Further, the mixed Nash equilibrium is not a strong equilibrium, since a coalition of size $N$, where one uploads and the others abstain, gives a higher payoff than that at mixed equilibrium.  

Combining these observations, we conclude that the set of strong equilibria consists exactly of all profiles where exactly one player uploads and all other players choose not to upload.
\end{proof}
\subsection{Proof of Corollary \ref{col1}}
\begin{proof} 
Recall that a pure strategy Nash equilibrium (NE) is an action profile $a^*$ such that no single player can unilaterally improve their payoff. A strong Nash 
equilibrium (SNE) is an action profile $a^*$ such that no coalition $S \subseteq \mathcal{N}$ can jointly deviate so that all members strictly improve their payoffs.
The set of pure NE is
\[
\text{Pure NE} = \{ a \in \{\text{U}, \text{NU}\}^N : m(a) =1 \}.
\]
where $b > c_f > c_l > 0$, $p>0$, and $m(a)$ is the number of uploaders in action profile $a$.  

Let player $i$ be the sole uploader: $a_i^* = \text{U}$ and $a_j^* = \text{NU}$ for all $j \neq i$. Then
\[
u_i(a^*) = b - c_f, \quad u_j(a^*) = b \text{ for } j \neq i.
\]
Define  $ W := \sum_{i \in \mathcal{N}} u_i(a)$. We have : 
\[
W(a^*)= Nb - c_f. 
\]
$W(a^*)$ is the maximum reward that could be distributed to the players. Hence, every pure NE is a global maximizer of $W$. Interestingly, this also implies that every pure NE is a strong equilibrium, as any socially optimal Nash equilibrium is a strong equilibrium \cite[Theorem~33]{SAMSON_CHAP1}. 

\end{proof}
\subsubsection{Proof of Proposition \ref{propess2}}
\begin{proof} 
Consider the symmetric game where all $N$ players belong to a single population. The expected payoff of a player using strategy $q$ against 
opponents using strategies $q_1, \dots, q_{N-1}$ is
$u(q, q_1, \dots, q_{N-1})$

For evolutionary stability analysis, we assume a monomorphic population where almost all players use $q_u^*$, except for a coalition of $k$ mutants using $q_m$.
Let $\epsilon = k/N$ denote the fraction of mutants. After deviation, the population profile is:
\[
\begin{cases}
\text{Fraction } \epsilon : \text{use } q_m,\\
\text{Fraction } 1-\epsilon : \text{use } q_u^*.
\end{cases}
\]

A resident player faces $N-1$ opponents, of which the number of mutants is distributed according to a hypergeometric-like distribution, reflecting $k$ mutants and $N-1-k$ residents.
 
Define
\[
D(q_m, \epsilon) = u(q_u^*, \epsilon, q_m) - u(q_m, \epsilon, q_m),
\]
where $u(q_u^*, \epsilon, q_m)$ and $u(q_m, \epsilon, q_m)$ denote the expected payoffs of a resident and a mutant, respectively.  

The resident strategy $q_u^*$ is evolutionarily stable against coalition size $k$ if $D(q_m, \epsilon) > 0$ for all $q_m \neq q_u^*$ and all $\epsilon \le k/N$.
 
Let $M$ denote the number of upload attempts among the $N-1$ opponents. Then:

\begin{itemize}
    \item For a resident: $M$ follows a binomial mixture reflecting $k$ mutants using $q_m$ and $N-1-k$ residents using $q_u^*$.
    \item For a mutant: $M$ follows a similar distribution with $k-1$ other mutants and $N-k$ residents.
\end{itemize}

The expected payoffs $u(q_u^*, \epsilon, q_m)$ and $u(q_m, \epsilon, q_m)$ can be computed exactly using the binomial distributions and the payoff structure of the game.

At $\epsilon = 0$, there are no mutants:
\[
D(q_m, 0) = u(q_u^*, q_u^*) - u(q_m, q_u^*).
\]

Since $q_u^*$ is a symmetric Nash equilibrium:
\[
u(q_u^*, q_u^*) \ge u(q_m, q_u^*) \quad \forall q_m.
\]

If $q_u^*$ is unique and the game is non-degenerate:
\[
D(q_m, 0) > 0 \quad \forall q_m \neq q_u^*.
\]

$D(q_m, \epsilon)$ is continuous in $\epsilon$ for fixed $q_m$, because the binomial probabilities vary continuously with population fractions and the payoff function is polynomial in probabilities.  

By continuity, for each $q_m \neq q_u^*$, there exists $\epsilon(q_m) > 0$ such that
\[
D(q_m, \epsilon) > 0 \quad \forall \epsilon < \epsilon(q_m).
\]

The invasion barrier $\alpha$ is
\[
\alpha = \min_{q_m \in (0,1),\, q_m \neq q_u^*} \epsilon(q_m) > 0.
\]

Equivalently, it can be characterized as
\[
\alpha = \max \left\{ \epsilon \in (0,1) : \min_{q_m \in (0,1),\, q_m \neq q_u^*} D(q_m, \epsilon) > 0 \right\}.
\]

By construction, any mutant fraction $\epsilon \le \alpha$ cannot invade the resident population.

\end{proof}

\end{document}